\documentclass[lettersize,journal]{IEEEtran}
\usepackage{bm}
\usepackage{amsmath,amsfonts}
\usepackage[ruled,linesnumbered]{algorithm2e}
\usepackage{array}
\usepackage[caption=false,font=scriptsize,labelfont=sf,textfont=sf]{subfig}
\usepackage{textcomp}
\usepackage{stfloats}
\usepackage{url}
\usepackage{blindtext}
\usepackage{hyperref}
\hypersetup{
	colorlinks=true,
	linkcolor=blue,
	filecolor=magenta,      
	urlcolor=cyan,
}
\usepackage{verbatim}
\usepackage{graphicx}
\usepackage{cite}
\usepackage{threeparttable}

\begin{document}
	\title{Traffic-Aware Hierarchical Beam Selection for Cell-Free Massive MIMO}
	
	\author{Chenyang Wang, Cheng Zhang,~\IEEEmembership{Member,~IEEE}, Fan Meng,\\ Yongming Huang,~\IEEEmembership{Senior Member,~IEEE}, Wei Zhang,~\IEEEmembership{Fellow,~IEEE}
		% <-this % stops a space
		\thanks{This work was supported in part by the National Natural Science Foundation of China under Grant 62271140, 62225107, and 62201394, and the Fundamental Research Funds for the Central Universities 2242022k60002. Part of this work has been accepted by the IEEE International Conference on Wireless Communications and Signal Processing (WCSP) 2023 \cite{10039343}. (Corresponding author: C. Zhang)}% <-this % stops a space
		\thanks{C. Wang, C. Zhang, and Y. Huang are with the National Mobile Communication Research Laboratory, the School of Information Science and Engineering, Southeast University, Nanjing 210096, China, and also with the Purple Mountain Laboratories, Nanjing 211111, China (e-mail: chenyang$\_$wang; zhangcheng$\_$seu; huangym@seu.edu.cn).}
		\thanks{F. Meng is with the Purple Mountain Laboratories, Nanjing 211111, China (e-mail: mengfan@pmlabs.com.cn).}
		\thanks{W. Zhang is with the School of Electrical Engineering and Telecommunications, the University of New South Wales, Sydney, NSW 2052, Australia, and also with the Purple Mountain Laboratories, Nanjing 211111, China (e-mail: wzhang@ee.unsw.edu.au).}}
	
	% The paper headers
	\markboth{SUBMITTED TO XXX}%
	{Shell \MakeLowercase{\textit{et al.}}: A Sample Article Using IEEEtran.cls for IEEE Journals}
	
	% Remember, if you use this you must call \IEEEpubidadjcol in the second
	% column for its text to clear the IEEEpubid mark.
	
	\maketitle
	
	\begin{abstract}
		Beam selection for joint transmission in cell-free massive multi-input multi-output systems faces the problem of extremely high training overhead and computational complexity.
		The traffic-aware quality of service additionally complicates the beam selection problem.
		To address this issue, we propose a traffic-aware hierarchical beam selection scheme performed in a dual timescale. In the long-timescale, the central processing unit collects wide beam responses from base stations (BSs) to predict the power profile in the narrow beam space with a convolutional neural network, based on which the cascaded multiple-BS beam space is carefully pruned. In the short-timescale, we introduce a centralized reinforcement learning (RL) algorithm to maximize the satisfaction rate of delay w.r.t. beam selection within multiple consecutive time slots. Moreover, we put forward three scalable distributed algorithms including hierarchical distributed Lyapunov optimization, fully distributed RL, and centralized training with decentralized execution of RL to achieve better scalability and better tradeoff between the performance and the execution signal overhead. Numerical results demonstrate that the proposed schemes significantly reduce both model training cost and beam training overhead and are easier to meet the user-specific delay requirement, compared to existing methods.
	\end{abstract}
	
	
	\section{Introduction}
	To realize low-latency, high-reliable, spectral- and energy-efficient communications in 5G and future 6G, the cell-free massive multi-input multi-output (CF-mMIMO) system enabled by distributed antenna system (DAS) and coordinated multi-point (CoMP) has been widely investigated in academic and industrial fields \cite{elhoushy2021cell}. In CF-mMIMO, a large number of BSs distributed in a small area are connected to a central processing unit (CPU) via fronthaul links for collaborative transmission. Compared to co-located MIMO systems, CF-mMIMO systems provide higher energy efficiency and fairer quality of service (QoS) for users within the coverage area due to the antenna deployment with greater flexibility \cite{zhang2019cell}. In the physical layer of CF-mMIMO systems, hybrid beamforming is a promising technique to further reduce the deployment cost and consuming energy, due to its lower radio frequency (RF) link requirements \cite{ahmed2018survey,zhang2018interleaved}.
	
	In practical systems, the analog design in hybrid beamforming generally involves codebook-based beam selection \cite{he2017codebook}.
	The beam selection enabled by the beam alignment/track mechanism in current 3GPP protocols, has large beam training overhead especially in mobile scenarios. Extensive studies have been developed to reduce the overhead and improve the efficient achievable rate~\cite{zhu2022beam,9573459,9422201,echigo2021deep,9175003,9832933,9410457,mengfan2022learning,9069211,zhang2020intelligent,10225317}. 
	In~\cite{zhu2022beam}, to obtain the optimal beam direction, a traceless Kalman filter is employed as a model-driven method to predict the user position or arrival angle at upcoming time slots. 
	In~\cite{9573459}, a direction-constrained compressed sensing based beam training scheme was proposed for the jitter effects on the mmWave channel response of unmanned aerial vehicles (UAV).
	
	Capitalizing on the powerful learning capabilities of machine learning (ML), ML-based beam selection schemes have been extensively studied~\cite{9422201,echigo2021deep,9175003,9832933,9410457,mengfan2022learning,9069211,zhang2020intelligent,10225317} to reduce the beam training overhead.
	%%\cite{9422201,echigo2021deep,9175003,9832933,9410457,mengfan2022learning,9069211,zhang2020intelligent,10225317}
	In~\cite{9422201}, the authors constructed an explicit mapping between transmit beams and physical coordinates via a Gaussian process and a reverse mapping that predicts physical coordinates from beam experiences via a hierarchical Bayesian learning model. This approach achieves a good beam alignment performance even with small sample sets.
	Deep learning (DL) incorporates formidable learning abilities to extract prior knowledge and make decisions in scenarios without explicit system models, e.g., learning the complicated relationship between channel state information at different beam-spaces~\cite{echigo2021deep}, frequency-spaces~\cite{9175003} and time-slots~\cite{9832933}, designing the beamforming to combat the propagation loss~\cite{9410457}. \cite{mengfan2022learning} transformed the high-dimensional beam prediction in the high-speed railway scenario into a two-stage task, i.e., a low-dimensional parameter estimation and a cascaded hybrid beamforming operation. 
	From the perspective of sensing the change in the environment, \cite{9069211} designed the action based on the beam index difference and formulated the problem of beam alignment as a stochastic bandit problem. 
	In addition, an environment interactive beam tracking framework was designed to realize model-and data-driven adaptive beam tracking, enabled by reinforcement learning (RL)~\cite{zhang2020intelligent}.
	Based on lightweight broad learning (BL), a user-side and a BS-side distributed incremental collaborative beam alignment approaches were proposed for CF-mMIMO downlink systems in~\cite{10225317}, which respectively realize implicit sharing of multiple user data and multiple BS features via reasonable distributed BL designs. However, these studies do not take the customized services w.r.t. the user QoS into consideration, leading to possible beam resource waste of the overall system. 
	
	Consequently, traffic-aware beam selection has become a hot research topic recently~\cite{fang2019queue, gatzianas2021traffic, venkatraman2016traffic, xu2021multi}. Studies in \cite{fang2019queue} and \cite{gatzianas2021traffic} considered stochastic traffic arrivals and transformed the problem of beam selection within multiple continuous time slots coupled to each other into a per-time-slot beam selection problem by incorporating Lyapunov optimization. Based on the modeling of the maximization problem under queue stability constraints, both the system delay and throughput can be improved.
	In \cite{venkatraman2016traffic}, an efficient beamforming design was proposed based on a small number of iterations and information exchanges to minimize the backlog queue. 
	In addition, an actor-critic RL approach is utilized to fulfill centralized training and distributed execution (CTDE) of beam selection in a collaborative cloud wireless edge network \cite{xu2021multi}, aiming to minimize the long-term average network latency with instantaneous QoS guarantee. 
	These studies assume equal arrival rates for multiple users while having no concerns about the user-specific requirements for differentiated QoS guarantee. Furthermore, either the beam selection for analog beamforming \cite{fang2019queue, gatzianas2021traffic, xu2021multi} rather than hybrid beamforming or the fully digital beamforming \cite{venkatraman2016traffic} is considered for CoMP coordinated beamforming (CB) where only channel information is exchanged among BSs and each user is only served by one BS.
	
	In this paper, we study the traffic-aware beam selection in CF-mMIMO systems with hybrid beamforming. Different from existing related works, we consider the joint transmission (JT) mode where multiple users are collaboratively served by multiple BSs. Moreover, the user-specific QoS requirement is considered. Specifically, the overall delay satisfaction rate of the system is adopted as the key performance indicator (KPI). The main contributions are summarized as follows:
	\begin{itemize}
		\item We propose a traffic-aware hierarchical beam selection scheme to solve the challenge of excessive cascaded beam space in CF-mMIMO hybrid beamforming. A long-timescale wide beam sweeping-based narrow beam prediction enabled by a convolutional neural network (CNN) helps reduce the feasible beam space for short-timescale traffic-aware beam selection, with negligible performance degradation and significant convergence acceleration.
		\item To maximize the overall delay satisfaction rate w. r. t. the pruned beam space, we model the traffic-aware beam selection problem as a partially observable Markov decision process (POMDP) and propose a dueling double deep $Q$ network (D3QN) based centralized scheme to improve the beam resource utilization. 
		
		\item To address the issues of high signaling exchange overhead and system delay in the centralized scheme for CF-mMIMO systems, we further propose three distributed schemes. First, we combine Lyapunov optimization and the proposed hierarchical framework to form the hierarchical distributed Lyapunov optimization (HDLO), which significantly improves the delay satisfaction rate compared to conventional Lyapunov optimization. 
		In addition, by treating the CF-mMIMO system as a multi-agent (MA) system, we enhance the original centralized scheme by employing two MA RL (MARL) paradigms~\cite{nguyen2020deep}, i.e., fully distributed and QMIX-based CTDE.

		\item Simulations show that 1) the proposed CNN-based narrow beam prediction can provide a good candidate beam set for following traffic-aware beam selection; 2) our proposed centralized hierarchical scheme achieves a higher delay satisfaction rate with a smaller model and beam training overhead compared to existing traffic-aware centralized schemes; 3) our proposed distributed schemes have acceptable performance degradation while reducing the signal overhead and accelerating the online execution speed, as compared to the centralized scheme.
	\end{itemize}
	
	The rest of the paper is organized as follows. Section~\ref{sec:sys} introduces the system model and problem formulation. The hierarchical beam selection framework and a centralized RL scheme for traffic-aware beam selection are proposed in Section~\ref{sec:cen}. Section~\ref{sec:dis} presents three distributed schemes based on the hierarchical beam selection framework. Section~\ref{sec:sim} provides experimental results of the proposed schemes. The paper is concluded in Section~\ref{sec:con}.
	
	Notations: bold upper case letters and lower case letters $\mathbf{A}(\mathbf{a})$ denote matrices and vectors, respectively. The conjugate transpose and transpose of $\mathbf{A}$ are denoted by $\mathbf{A}^{\mathsf{H}}$ and $\mathbf{A}^{\mathsf{T}}$. The Kronecker product is represented by $\otimes$. $\text{blkdiag}(\cdot)$ is the operator for the block diagonal matrix. $\mathcal{CN}(0,1)$ represents the circularly symmetric complex Gaussian distribution with mean 0 and variance 1. $\mathbf{I}_N$ denotes the $N$-dimensional eye matrix.

	\section{System Model and Problem Formulation}\label{sec:sys}
	We consider a CF-mMIMO system with $B$ BSs serving $U$ single-antenna users simultaneously. Each BS is equipped with $M$ antennas fully connected to $U$ RF chains, and all BSs are connected to the CPU via the fronthaul links. Define $\mathbb{U}=\{1,...,U\}$ and  $\mathbb{B}=\{1,...,B\}$.
	\subsection{Channel Model}
	According to the Saleh-Valenzuela model, the channel from user $u$ to BS $b$ at time slot $t$ can be expressed as
	\begin{equation}
		\mathbf{h}_{b,u}(t)=\sqrt{\frac{M}{L(t)}}\sum_{l=1}^{L(t)}g_{b,u,l}(t)\mathbf{a}\left(\theta_{b,u,l}(t),\phi_{b,u,l}(t)\right),
	\end{equation}
	where $L$ is the number of propagation paths and $l$ is the path index.
	$\theta$ and $\phi$ are the azimuth and elevation angles. $g=\alpha\beta$ is the complex gain where $\alpha$ and  $\beta$  are the large-scale and the small-scale fadings respectively. Considering a uniform planar array (UPA) with half-wavelength antenna spacing, the array response vector is $\mathbf{a}\left(\theta_{b,u,l},\phi_{b,u,l}\right)=\mathbf{a}_{z}(\phi_{b,u,l})\otimes\mathbf{a}_{y}(\theta_{b,u,l},\phi_{b,u,l})$ where $\mathbf{a}_z\left(\phi_{b,u,l}\right)=\frac{1}{\sqrt{M_z}}\left[1,...,e^{\jmath \pi(M_z-1)\cos\phi_{b,u,l}}\right]^{\mathsf{T}}$ and $\mathbf{a}_y\left(\theta_{b,u,l},\phi_{b,u,l}\right)=\frac{1}{\sqrt{M_y}}\left[1,...,e^{\jmath\pi(M_y-1)\sin\theta_{b,u,l}\sin\phi_{b,u,l}}\right]^{\mathsf{T}}$ respectively represent the vertical and horizontal antenna array responses. The numbers of antennas in the horizontal and vertical directions are $M_y$ and $M_z$, and $M=M_yM_z$. 
	
	We consider a time-varying channel model with a dual timescale. The large-scale fading $\alpha$, the azimuth angle $\theta$, the elevation angle $\phi$ and the number of paths $L$ change only between long-timescale intervals that contain multiple time slots, while the small-scale fading $\beta$ changes across time slots~\cite{xu2004generalized}.
	
	\subsection{Transmission Model}
	%Here the time slot $t$ denotes the continuous time period $[t\tau,(t+1)\tau)$ where $\tau$ denotes the transmission time slot of the system. 
	In each time slot, beam training and data transmission are sequentially conducted. The channel from user $u$ to all BSs is defined as $\mathbf{h}_u=\left[\mathbf{h}_{1,u}^\mathsf{T},\mathbf{h}_{2,u}^\mathsf{T},...,\mathbf{h}_{B,u}^\mathsf{T}\right]^\mathsf{T}\in \mathbb{C}^{BM\times 1}$, and the channel state information (CSI) of the system is $\mathbf{H}=[\mathbf{h}_1,\mathbf{h}_2,...,\mathbf{h}_U]\in \mathbb{C}^{BM\times U}$. The BS selects the analog beam from the Discrete Fourier Transform (DFT) codebook $\mathbf{F}\in \mathbb{C}^{M\times M}$. The index of the $u$-th analog beam selected by the BS $b$ is denoted by $i_{b,u}$. Defining the analog combiner $\mathbf{W}_{\text{RF},b}=\left[\mathbf{w}_{\text{RF},b,1},\mathbf{w}_{\text{RF},b,2},...,\mathbf{w}_{\text{RF},b,U}\right]^{\mathsf{H}}\in \mathbb{C}^{U\times M}$ for BS $b$ and the analog combiner $\mathbf{W}_{\text{RF}}=\text{blkdiag}(\mathbf{W}_{\text{RF},1},\mathbf{W}_{\text{RF},2},...,\mathbf{W}_{\text{RF},B})\in \mathbb{C}^{BU\times BM}$ for all BSs, the equivalent CSI is $\overline{\mathbf{H}}=\mathbf{W}_{\text{RF}}\mathbf{H}\in \mathbb{C}^{BU\times U}$, and the CPU designs the digital zero-forcing combiner by $\mathbf{W}_{\text{BB}}=\left(\overline{\mathbf{H}}^\mathsf{H}\overline{\mathbf{H}}\right)^{-1}\overline{\mathbf{H}}^\mathsf{H}\in \mathbb{C}^{U\times BU}$. After combining, the signal of user $u$ is 
	\begin{equation}
		\hat{s}_u=\sqrt{P_u}\mathbf{w}_u^\mathsf{T}\mathbf{h}_us_u+\sum_{v\neq u}\sqrt{P_v}\mathbf{w}_u^\mathsf{T}\mathbf{h}_vs_v+\mathbf{w}_u^\mathsf{T}\mathbf{n},
	\end{equation}
	where $P_u$ denotes the average normalized transmit power, $\mathbf{w}_u^\mathsf{T}$ is the $u$-th row of the hybrid combiner $\mathbf{W}=\mathbf{W}_{\text{BB}}\mathbf{W}_{\text{RF}}\in \mathbb{C}^{U\times BM}$, $s_u$ denotes the baseband signal satisfying $\mathbb{E}\{|s_u|^2\}=1$,  $\mathbf{n}=[\mathbf{n}_1^\mathsf{T},\mathbf{n}_2^\mathsf{T},...,\mathbf{n}_B^\mathsf{T}]^\mathsf{T}$ where $\mathbf{n}_b\in \mathcal C\mathcal N(\mathbf{0},\mathbf{I}_M)$ denotes the receiver noise of BS $b$. The SINR expression for user $u$ is given by
	\begin{equation}
		\rho_{u}=\frac{P_u\Vert\mathbf{w}_u^\mathsf{T}\mathbf{h}_{u}\Vert^2}{\sum_{v\neq u}P_v\Vert\mathbf{w}_{u}^\mathsf{T}\mathbf{h}_{v}\Vert^2+\Vert\mathbf{w}_{u}\Vert^2}.
	\end{equation}
	Thus, the achievable rate of user $u$ can be obtained as 
	\begin{equation}\label{equ:actual_rate}
		R_u = W_u\frac{\tau-N_{\text{tr}}\tau_\text{c}}{\tau}\log_2\left(1+\rho_u\right),
	\end{equation}
	where $W_u$ denotes the available bandwidth, $N_{\text{tr}}$ is the symbol number for beam training, $\tau$ denotes the time slot duration and $\tau_\text{c}$ is the symbol duration. 
	
	Assuming that the length of newly arrived packets follows the Pareto distribution with shape $\kappa$ and threshold $\chi_{\min}$~\cite{796159}. $\chi_u^i$ denotes the $i$-th newly arrived packet of user $u$ and $|\chi_u^i|$ is the counterpart packet length. The probability density function (PDF) of $|\chi_u^i|$ can be expressed as
	\begin{equation}
		p\left(\left|\chi_u^i\right|\right)=\begin{cases}
			0,& \left|\chi_u^i\right|<\chi_{\min},\\
			\frac{\kappa \chi_{\min}}{\left|\chi_u^i\right|^{\kappa+1}}, & \left|\chi_u^i\right|>\chi_{\min}.
		\end{cases}
	\end{equation}
	We assume that the new packets only arrive at the end of each time slot, and the number of newly arriving packets within a time slot follows the Poisson distribution \cite{leung1994traffic} as follows
	\begin{equation}
		{p}\left(I_u(t)=n\right) = \frac{\lambda_u^{n}}{n!}\exp(-\lambda_u),
	\end{equation}
	where $\lambda_u$ denotes the mean of $I_u(t)$. Finally, the new arrival traffic of user $u$ at time slot $t$ can be defined as
	\begin{equation}
		A_u(t)=\sum_{i=1}^{I_u(t)}\left|\chi_u^i\right|.
	\end{equation}
	Accordingly, the traffic that can be served by each user in time slot $t$ is $\tau R_u(t)$. For user $ u $, we define $q_u(t)$ as the traffic to be processed at the beginning of time slot $t$, then the traffic to be processed at the beginning of next time slot $t+1$ is
	\begin{equation}\label{equ:queue}
		q_u(t+1) = \max\{q_u(t)-\tau R_u(t),0\}+A_u(t).
	\end{equation}

	\subsection{Problem Formulation}
	During the entire period of uploaded traffic which contains $T$ time slots, maximizing the overall delay satisfaction rate of the system w.r.t. the beam selections can be formulated as follows:
	\begin{eqnarray}\label{equ:org}
		\underset{b\in \mathbb{B},u\in \mathbb{U}}{\underset{i_{b,u}(1),...,i_{b,u}\left(T\right)}{\max}} \sum_{u=1}^U \text{Pr}(\tilde{d}_u<\bar{d}_u),
	\end{eqnarray}
	where $\tilde{d}_u$ and $\bar{d}_u$ denote the average delay and delay requirement of user $u$, respectively. The average queue length of user $u$ is
	\begin{equation}
		\tilde{q}_u=\dfrac{1}{T}\sum_{t=2}^{T+1}q_u(t).
	\end{equation}
	In queueing theory \cite{gross2008fundamentals}, Little's theorem states that the average queue length $\tilde{q}_u=\omega_u \tilde{d}_u$ in the steady state of the system, where $\omega_u$ denotes the average arrival rate of traffic.
	Letting the queue length requirement $\bar{q}_u=\omega_u\bar{d}_u$, the problem~\eqref{equ:org} can be equivalently transformed as
	\begin{eqnarray}\label{equ:max}
		\underset{b\in \mathbb{B},u\in \mathbb{U}}{\underset{i_{b,u}(1),...,i_{b,u}\left(T\right)}{\max}} \sum_{u=1}^U\text{Pr}(\tilde{q}_u<\bar{q}_u).
	\end{eqnarray}
	
	\section{Hierarchical Centralized Beam Selection}\label{sec:cen}
	\begin{figure}[tbp]
		\centering
		\includegraphics[width=0.74\linewidth]{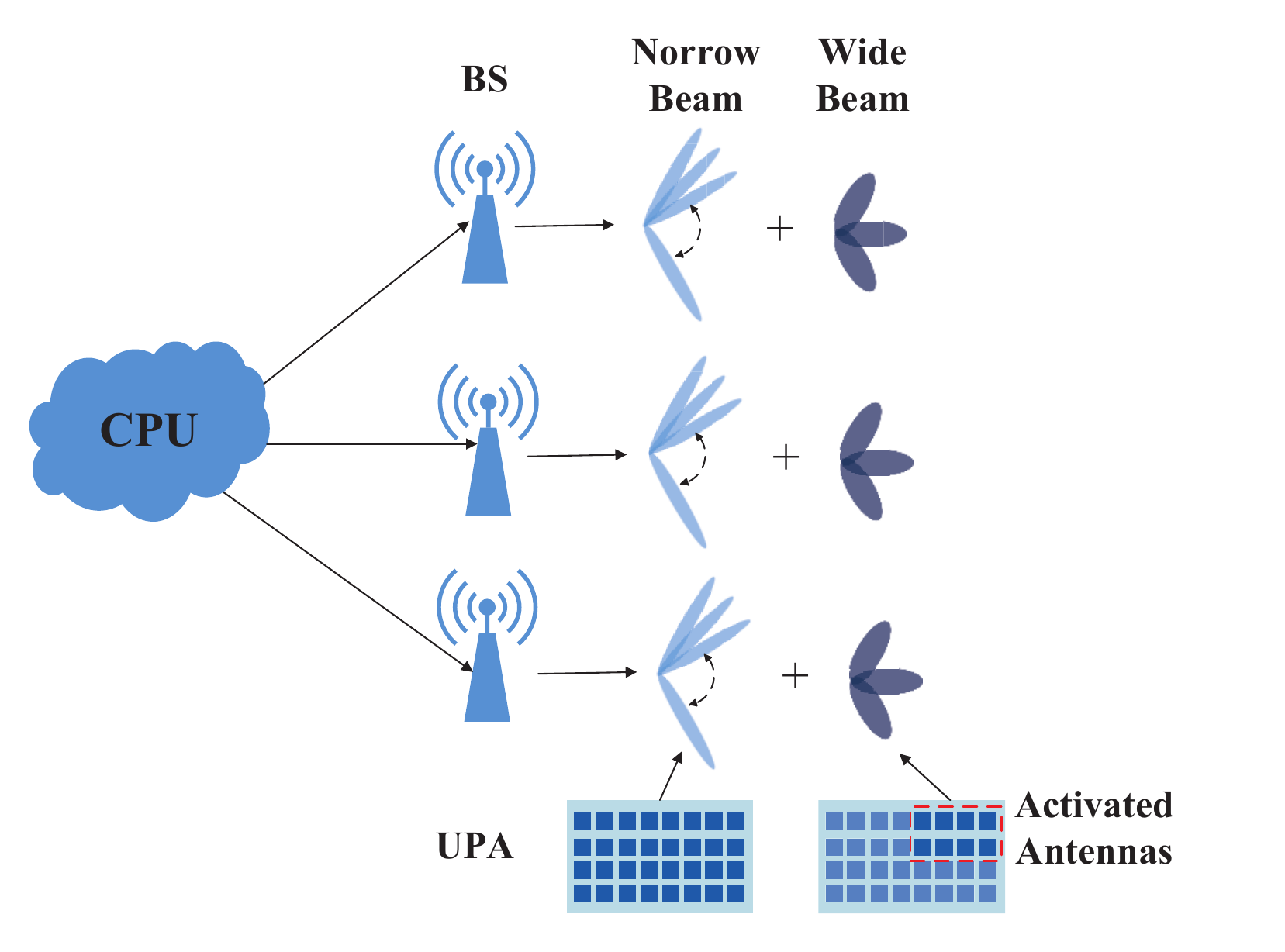}
		\caption{Wide beam based narrow beam prediction.}
		\label{pic:broadbeam}
	\end{figure}

	Considering problem~\eqref{equ:max}, the action space of CF-mMIMO systems grows exponentially with the numbers of cooperating BSs and users, which makes it difficult for RL to converge and adapt to the fast time-varying channels in real-time. As illustrated in Fig.~\ref{pic:broadbeam}, by exploiting the correlation between the signal strengths of wide and narrow beams in angle, we propose an action space pruning method based on wide-beam responses to accelerate the convergence of RL and reduce the computational cost of training. The wide beam is formed by activating a small number of antennas using the standard DFT codebook.
	
\begin{figure}[tbp!]
		\centering 
		\includegraphics[width=0.9\linewidth]{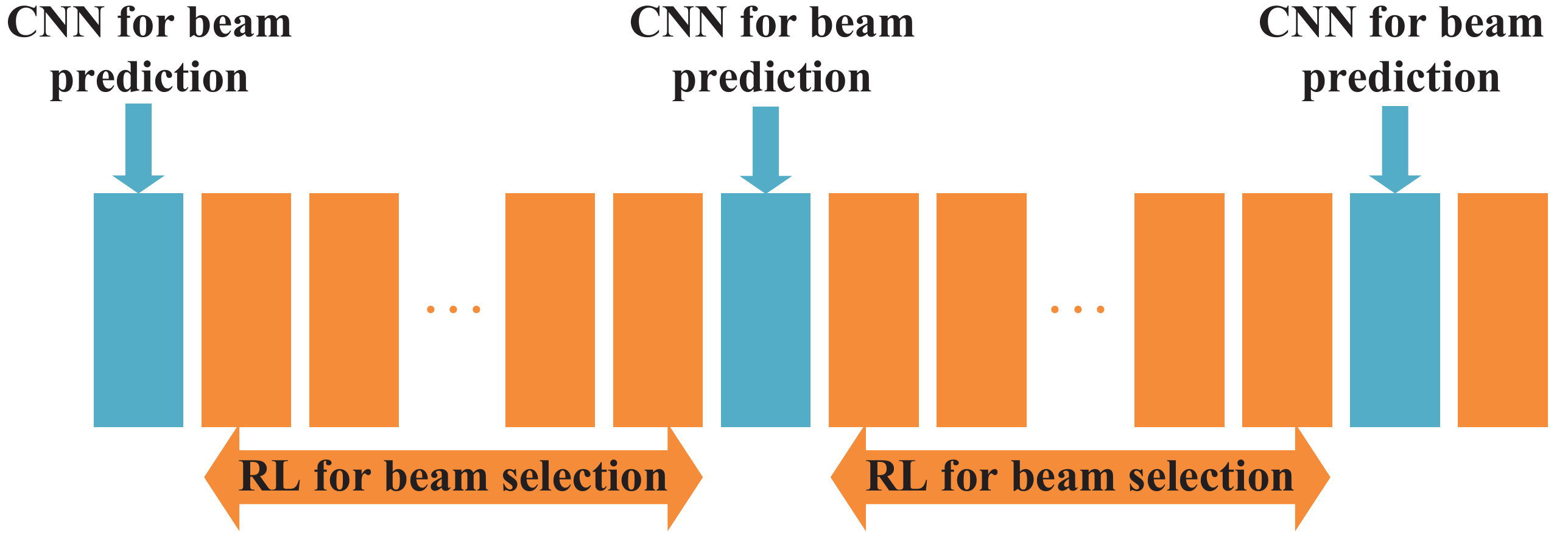} 
		\caption{The illustration of the proposed hierarchical beam selection.}
		\label{pic:frame}
	\end{figure}

	As shown in Fig.~\ref{pic:frame}, the proposed hierarchical beam selection is two-fold. Firstly in the long-timescale interval, the BSs receive reference signals from one user with wide beams, then the CPU aggregates the measured wide beams to infer the narrow beam power profile and selects a candidate narrow beam set by predicted power. Secondly in the short-timescale, the candidate narrow beam sets are aggregated as an action space, and we propose a centralized RL algorithm to solve problem~\eqref{equ:max}. Specifically, the CPU selects beams for the users in the aggregated action space based on the current system state, followed by data upload.
	\subsection{Wide Beam Response Based Action Space Pruning}
	Motivated by the previous research~\cite{alkhateeb2018deep} that utilizes omni-directional sensing beams from multiple BSs to predict the effective achievable rates of narrow beams for each BS, we propose a predictive method for the narrow beam responses of the collaborative BSs based on multiple wide beams of the collaborative BSs. In contrast to the approach described in~\cite{alkhateeb2018deep}, which can only be executed in the CPU, the proposed method can also be executed locally in each BS, i.e., each BS utilizes the local multiple wide beam responses to predict its narrow beam responses. Furthermore, while previous research primarily concentrates on the accuracy of predicting the strongest narrow beam, the introduced paradigm, based on the wide-beam response at the beginning of the long-timescale interval, primarily serves to provide a compressed action space for the dynamic planning problem of beam selection oriented to maximize the delay satisfaction rate in each time slot during the long-timescale interval. Predicting the strongest beam only is found to be insufficient in yielding optimal performance, attributing to the presence of multi-user interference and the varying traffic demands. Hence, the effectiveness of the proposed method depends on the ability to accurately predict a set of narrow beams with relatively high strength, denoted as the \emph{candidate beam set}.
	
	To obtain a good candidate beam set, we propose a CNN-based beam predictor to learn the relationship between the strength of the wide beam and that of the narrow beam. Specifically, the prediction is conducted user-by-user. Considering user $ u $, the CPU receives the wide beam strengths of all BSs, i.e., $ \{\bm{\eta}_{\textup{w}, b, u}\}_{b=1}^B $, and uses a CNN to predict the corresponding narrow beam strengths, i.e., $ \{\bm{\eta}_{\textup{n}, b, u}\}_{b=1}^B $, where the subscripts $ (\cdot)_{\textup{w}} $ and $ (\cdot)_{\textup{n}} $ respectively denote the wide and narrow beams. The prediction is modeled as a regression problem and we use mean square error (MSE) as the cost function. The parameters of the CNN are iteratively updated by mini-batch gradient descent (MBGD) until convergence.
	
	The received signal strength $ \eta $ is significantly affected by the distance between the user and the BS. Inspired by \cite{alkhateeb2018deep}, we propose a per-BS normalization to tackle this issue, where all the beam strengths from a user to a BS are normalized by dividing the maximum beam strength. The normalized beam strengths of the BS $b$ serving user $u$ is:
	\begin{equation}\label{norm}
		\bm{\eta}_{b,u}'=\frac{\bm{\eta}_{b,u}}{\max_{j}\eta_{b,u,j}}.
	\end{equation} 
	The normalization is applied to the input and also the output. In this study, the proposed per-BS normalization is experimentally found to outperform per-dataset, per-sample and per-element methods, in terms of the prediction accuracy of the candidate beam set.
	
	In the CF system, the probability of beam conflicts between users greatly increases as the user number increases. The existing work that only selects the strongest predicted narrow beam, fails to address this problem. To handle beam conflicts and also ensure QoS, we construct the candidate beam set that includes several strongest predicted narrow beams. 
	%\textcolor{red}{However, enlarging the candidate beam set only gives the opportunity to avoid beam conflict, which occurs when the beams are not appropriately selected.} 
	However, even if the candidate beam set is large, beam conflicts may still occur, when the beams are not appropriately selected.
	To further avoid beam conflicts for users, we propose three-fold action space pruning to remove the duplicated and conflicting actions and form an effective action space $\mathcal{A}_{\text{eff}}$. First, after conflicting action removal, the action space of the BS $b$ is pruned as 
	\begin{align}\label{equ:eff_b1}
		\mathcal{A}_{b} =& \left\{\mathbf{a}_{b} =  \{\mathbf{f}_u\}_{u=1}^U | \mathbf{f}_u \in \mathcal{C}_{b,u}, \mathbf{f}_u \neq \mathbf{f}_{u'}, \right.\nonumber
		\\ &\hspace{3cm} \left. 
		\forall u, u' \in \mathbb{U}, u' \neq u \right\},
	\end{align}
	%		\begin{equation}\label{equ:eff_b1}\footnotesize
		%		\mathcal{A}_{b} = \left\{\mathbf{a}_{b} =  \{\mathbf{f}_u\}_{u=1}^U |\right.\\\left. \mathbf{f}_u \in \mathcal{C}_{b,u}, \mathbf{f}_u \neq \mathbf{f}_{u'}, \forall u, u' \in \mathbb{U}, u' \neq u \right\},
		%	\end{equation}
	where $\mathcal{C}_{b,u}$ denotes the candidate beam set for user $u$ provided by BS $b$. Second after duplicated action removal, the effective action space of BS $ b $ is 
	\begin{equation}\label{equ:eff_b}
		\mathcal{A}_{\text{eff}, b} = \left\{\mathbf{a}_{b} | \mathbf{a}_{b} \neq \mathbf{a}'_{b}, \forall \mathbf{a}_{b}, \mathbf{a}'_{b} \in \mathcal{A}_{b}\right\}.
	\end{equation}
	Third, the effective action spaces of all BSs is a cartesian product of $ \left\{\mathcal{A}_{\text{eff}, b}\right\}_{b=1}^B $, i.e.,  $\mathcal{A}_{\text{eff}}=\mathcal{A}_{\text{eff},1}\times\cdots\times\mathcal{A}_{\text{eff},B}$. 
	
	\begin{figure}[tbp]
		\centering
		\includegraphics[width=0.95\linewidth]{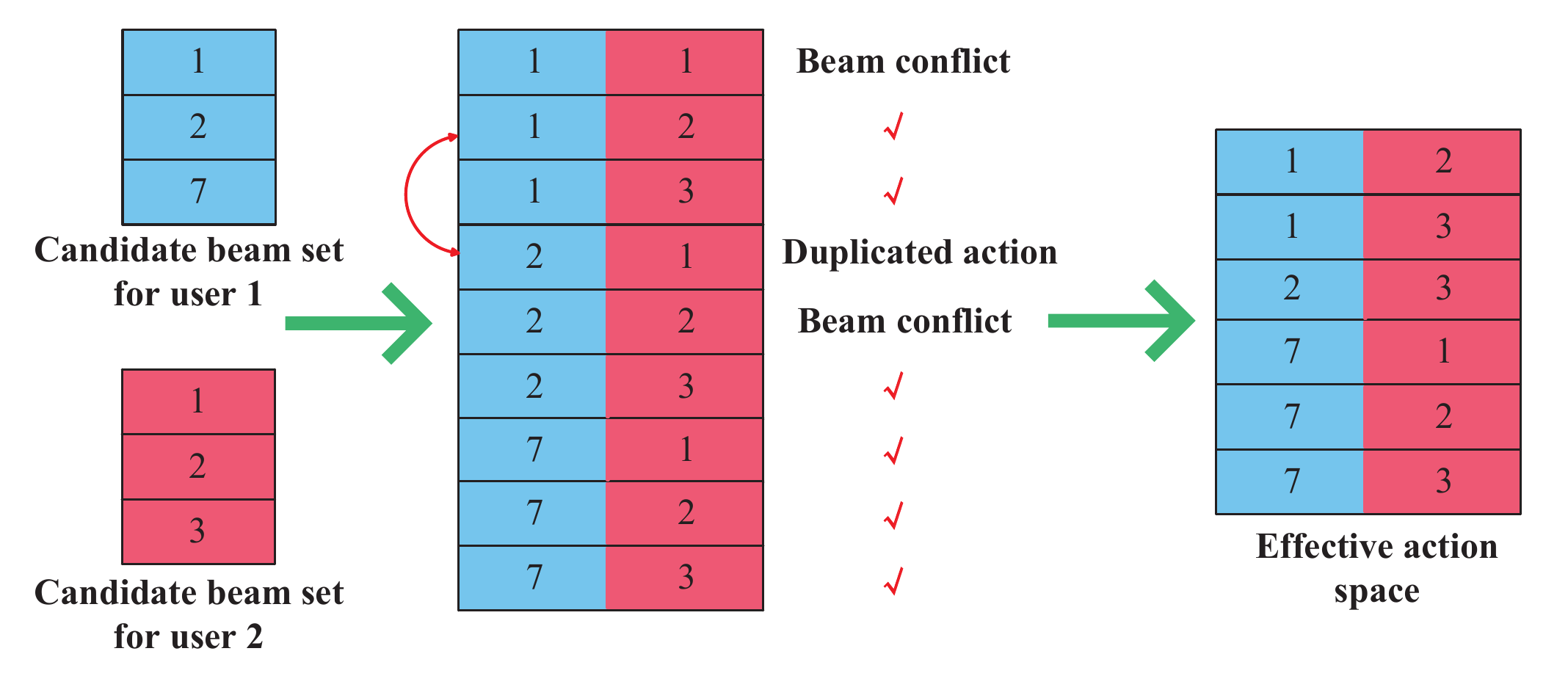}
		\caption{Remove the duplicated and conflicting actions and obtain the effective action space (Two-users case).}
		\label{pic:action_selection}
	\end{figure}
	
	In Fig.~\ref{pic:action_selection}, we take two users as an example to illustrate the proposed pruning. The candidate beam sets provided by BS $b$ to user 1 and user 2 respectively are $\mathcal{C}_{b,1}=\{\mathbf{f}_1,\mathbf{f}_2,\mathbf{f}_7\}$ and $\mathcal{C}_{b,2}=\{\mathbf{f}_1,\mathbf{f}_2,\mathbf{f}_3\}$. The primary action space is $ \mathcal{C}_{b,1}\times\mathcal{C}_{b,2}=\left\{\{\mathbf{f}_1,\mathbf{f}_1\},\{\mathbf{f}_1,\mathbf{f}_2\},\{\mathbf{f}_1,\mathbf{f}_3\},\{\mathbf{f}_2,\mathbf{f}_1\},\{\mathbf{f}_2,\mathbf{f}_2\},\{\mathbf{f}_2,\mathbf{f}_3\},\{\mathbf{f}_7,\mathbf{f}_1\}, \right.\\ \left.\\\{\mathbf{f}_7,\mathbf{f}_2\},\{\mathbf{f}_7,\mathbf{f}_3\}\right\}$, resulting from the intersection of the two candidate beam sets. However, it is observed that there exists a beam conflict in action $\{\mathbf{f}_1,\mathbf{f}_1\}$ and action $\{\mathbf{f}_2,\mathbf{f}_2\}$, while action $\{\mathbf{f}_2,\mathbf{f}_1\}$ and action $\{\mathbf{f}_1,\mathbf{f}_2\}$ are repeated, they are excluded from the actual action space of the BS $b$. The effective action space of the BS $b$ $\mathcal{A}_{\text{eff},b}$ is obtained after the exclusion. 
	
	Notably, the size of the effective action space is substantially smaller compared to that of the original and actual action space. The reduction in this action space size will accelerate the convergence speed of RL and improve the delay satisfaction rate during the training phase.
	
	\subsection{Centralized RL Algorithm for Problem~\eqref{equ:max}}
	
	In the centralized processing for a CF system, the CPU (i.e., the agent) interacts with the environment to optimize the delay satisfaction rate via multiple BSs connected to it. Due to the unavailability of a global state, the problem~\eqref{equ:max} is formulated as a POMDP, which is modeled as follows:
	
	\subsubsection{\textbf{State and Observation}}
	
	The state $\mathbf{s}(t)$ is the concatenation of the amount of traffic to be processed and the CSI of all users, i.e. $\mathbf{s}(t)=\{q_u(t),\mathbf{H}(t),u\in\mathbb{U}\}$.
	The probability of state $\mathbf{s}(t)$ moving to state $\mathbf{s}(t+1)$ after taking action $\mathbf{a}(t)$ is expressed as
	\begin{align}
		\nonumber
		&\text{Pr}\left(\mathbf{s}(t+1)|\mathbf{s}(t),\mathbf{a}(t)\right)\\\nonumber
		&=\text{Pr}\left(\mathbf{H}(t+1),\mathbf{q}(t+1)|\mathbf{s}(t),\mathbf{a}(t)\right)\\\nonumber
		&=\text{Pr}\left(\mathbf{H}(t+1)|\mathbf{s}(t),\mathbf{a}(t)\right)\text{Pr}\left(\mathbf{q}(t+1)|\mathbf{s}(t),\mathbf{a}(t)\right)\\ 	
		&=\text{Pr}\left(\mathbf{H}(t+1)|\mathbf{H}(t)\right)\prod_{u=1}^{U}\text{Pr}\left(q_u(t+1)|\mathbf{s}(t),\mathbf{a}(t)\right),
	\end{align}    
	where $\mathbf{q}(t)=\{q_1(t),q_2(t),...,q_u(t)\}$ denotes the set of all user queue lengths in the system.

	Given the high overhead of full CSI, 
	a more realistic assumption is that the CPU only acquires the equivalent CSI.
	The observation at time slot $t$ can then be expressed as
	\begin{equation}\label{observation_eq}
		\mathbf{o}(t)=\{\mathbf{q}(t+1),\overline{\mathbf{H}}(t)\}.
	\end{equation}
	
	\subsubsection{\textbf{Action}}
	
	The action is denoted as the set of beam indexes assigned by each BS to its users, i.e., $\mathbf{a}(t)=\{i_{b,u}(t),b\in\mathbb{B},u\in\mathbb{U}\}$.
	
	\subsubsection{\textbf{Reward}}
	Problem~\eqref{equ:max} is intrinsically tied to considerations involving both the mean $\mathbb{E}\{\tilde{q}_u\}$ and the variance ${\rm{Var}}\{\tilde{q}_u\}=\mathbb{E}\{\left(\tilde{q}_u-\mathbb{E}\left\{\tilde{q}_u\right\}\right)^2\}$. Specifically, the goal can be set to minimize both  $\mathbb{E}\{\tilde{q}_u\}$ and ${\rm{Var}}\{\tilde{q}_u\}$.
	In a stationary environment, acquiring more accurate estimates for $\mathbb{E}\{\tilde{q}_u\}$ and ${\rm{Var}}\{\tilde{q}_u\}$ demands the completion of several episodes. Besides, it will cause serious estimation errors when we use the previous queue observations to estimate the mean and variance. Moreover, the PDF of the reward would vary with $\mathbb{E}\{\tilde{q}_u\}$ and ${\rm{Var}}\{\tilde{q}_u\}$, which is more difficult for the CPU to learn. To overcome these issues, the risk is utilized~\cite{garcia2015comprehensive} which generally indicates that the system will enter a fault state. In this work, fault state is defined as the instantaneous queue length $q_u$ exceeding the queue threshold $\breve{q}_u$. 
	To maximize the delay satisfaction rate, we propose to incorporate risk-sensitive safe RL~\cite{garcia2015comprehensive} and design the reward as follows
	\begin{equation}\label{equ:cost}
		r(t)=-\sum_{u=1}^U \dfrac{q_{u}(t+1)}{\bar{q}_u}-\delta\sum_{u=1}^{U} \mathbb{I}(q_u(t+1)>\breve{q}_u).
	\end{equation}
	The first part of the reward is to allocate more resources for users with high delay requirements. The second part denotes the risk that restricts the users' queue lengths, and $\delta$ is the weighting factor.
	
	According to the POMDP theory, it is possible to construct a pseudo-state using the observations, actions and rewards up to time slot $t$ as sufficient statistics~\cite{seo2015training} as follows
	\begin{equation}\label{equ:state}
		\hat{\mathbf{s}}(t)=\{\mathbf{o}(t-1),\mathbf{a}(t-1),r(t-1),...,\mathbf{o}(1),\mathbf{a}(1),r(1)\}.
	\end{equation}
	Due to the temporal correlation between consecutive time slots, a small number of previous time slots may be sufficient to construct the pseudo-state~\cite{park2022deep}.

	Since the state transition probabilities are unknown, the model-free RL is an effective way to solve MDP problems. Considering that the pruned action space is still large, D3QN can learn the difference between actions, which is very effective for the environment with a large action space. Therefore, we solve the POMDP problem with D3QN. 
	The D3QN utilizes the current pseudo-state of the system as its input $\hat{\mathbf{s}}$. Subsequently, this input undergoes forward propagation through the neural network to yield the action-state value function $Q(\hat{\mathbf{s}},\mathbf{a})$ for each action within the effective action space. The action is selected using $\epsilon$-greedy exploration during the training phase
	\begin{align}\label{equ:explore}
		\mathbf{a}=\begin{cases}
			\arg\max_{\mathbf{a}\in\mathcal{A}_{\text{eff}}}{Q}(\hat{\mathbf{s}},\mathbf{a}), & 1-\epsilon\\
			\mathbf{a}\in\mathcal{A}_{\text{eff}}, & \epsilon
		\end{cases}.
	\end{align}
	While the action with the highest $Q$ value is selected during the execution phase.

	In D3QN, the model branches into two subnetworks after feature extraction: the first, called the value network, is responsible for estimating state values through a single output node; and the second, called the advantage network, computes the advantage value of taking a specific action in the current state. Mathematically, the state-action value function $Q(\hat{\mathbf{s}},\mathbf{a})$ can be obtained as
	\begin{equation}\label{equ:cal_q}
		Q(\hat{\mathbf{s}},\mathbf{a})=V(\hat{\mathbf{s}})+A(\hat{\mathbf{s}},\mathbf{a})-\frac{1}{\left|\mathcal{A}_{\text{eff}}\right|}\sum_{\mathbf{a}'\in\mathcal{A}_{\text{eff}}}A(\hat{\mathbf{s}},\mathbf{a}'),
	\end{equation}
	where $V(\cdot)$ denotes the state value function, $A(\cdot)$ denotes the advantage function. The network $Q_\phi$ is updated using the target network $Q_{\phi^-}$ to mitigate overestimation of the $Q$ value. The target network $Q_{\phi^-}$ is initialized to be the same as the network $Q_\phi$ and updated every $N_c$ steps with the current $Q$ network's parameters $\phi^-\leftarrow\phi$. The loss function for training the $Q$ network can be expressed as follows
	%	\begin{equation}\label{equ:loss_tot}\scriptsize
		%			L(\phi)=\sum_{n=1}^{N}\big(r_n+\gamma Q_{\phi^-}(\hat{\mathbf{s}}'_n,\underset{\mathbf{a}_n'\in\mathcal{A}_{\text{eff}}}{\arg\max}Q_\phi(\hat{\mathbf{s}}'_n,\mathbf{a}'_n))
		%			-Q_\phi(\hat{\mathbf{s}}_n,\mathbf{a}_n)\big)^2,
		%		\end{equation}
	%{\color{red}
		%			\begin{align}\label{equ:loss_tot}
			%			L(\phi)=&\sum_{n=1}^{N}\left(r_n+\gamma Q_{\phi^-}\left(\hat{\mathbf{s}}'_n,\underset{\mathbf{a}_n'\in\mathcal{A}_{\text{eff}}}{\arg\max}Q_\phi\left(\hat{\mathbf{s}}'_n,\mathbf{a}'_n\right)\right)\right. \nonumber \\  
			%			&\hspace{4.3cm}\left.
			%			-Q_\phi(\hat{\mathbf{s}}_n,\mathbf{a}_n)\right)^2,
			%		\end{align}}
		\begin{align}\label{equ:loss_tot}
			L(\phi)=&\sum_{n=1}^{N}\bigg(r_n+\gamma Q_{\phi^-}\Big(\hat{\mathbf{s}}'_n,\underset{\mathbf{a}_n'\in\mathcal{A}_{\text{eff}}}{\arg\max}Q_\phi\left(\hat{\mathbf{s}}'_n,\mathbf{a}'_n\right)\Big) \nonumber \\  
			&\hspace{4.1cm}
			-Q_\phi(\hat{\mathbf{s}}_n,\mathbf{a}_n)\bigg)^2,
	\end{align}where $N$ denotes the batch size sampled from the replay buffer, $\gamma$ denotes the discount factor, $\hat{\mathbf{s}}'$ denotes the next state after taking action $\mathbf{a}$ in the current state $\hat{\mathbf{s}}$.
	
	At the beginning of each long-timescale interval, the effective action space $\mathcal{A}_{\text{eff}}$ is determined by the wide beam response. In each time slot, the CPU determines the beam set $\mathbf{a}$ to optimize the delay satisfaction rate, then observes the change in users' queue lengths $\mathbf{q}$ and records this state transition $<\hat{\mathbf{s}},\mathbf{a},r,\hat{\mathbf{s}}'>$ in the replay buffer. MBGD is utilized to train the D3QN. The proposed wide beam response-based D3QN (WBR-D3QN) is summarized in Algorithm~\ref{alg:WBR-D3QN}.

	\begin{algorithm}[ht!]
		\caption{WBR-D3QN (Training phase)}
		\label{alg:WBR-D3QN}
		
		\BlankLine
		All BSs receive wide beam response and the CPU constructs the effective action space by the prediction of CNN.
		
		Initialize the replay buffer $\mathcal{D}$, the $Q$ network with random parameters $\phi$, and the target $Q$ network with parameters $\phi^-=\phi$.
		
		\BlankLine
		\For{$episode=1:total\_episode$}{
			Reset the environment, and randomly initialize queues for all users.
			
			\For{$t=1:T$}{
				\eIf{$t=1$}{
					The CPU randomly chooses the beam set $\mathbf{a}(t)$ and passes the beam indices to each BS.
				}{The CPU determines the beam set $\mathbf{a}(t)$ according to Eq.~\eqref{equ:explore} and passes the beam indices to each BS for the pseudo-state $\hat{\mathbf{s}}(t)$.}
				% 随机选择一个动作  
				% else
				
				% 根据伪状态选择动作
				
				% 每个基站执行波束动作，并在时隙末统计获得的奖励值和新伪状态。
				After the BSs conduct the beam selection and data transmission, the CPU collects the equivalent CSI $\overline{\mathbf{H}}(t)$ and the queue length $\mathbf{q}(t+1)$ to constrcut the new pseudo-state $\hat{\mathbf{s}}(t+1)$ in Eq.~\eqref{equ:state}.  
				
				The CPU stores the current transition $<\hat{\mathbf{s}}(t),\mathbf{a}(t),r(t),\hat{\mathbf{s}}(t+1)>$ in $\mathcal{D}$.
				
				The CPU updates the $Q_\phi$ network according to Eq.~\eqref{equ:loss_tot}.
				
				After $N_c$ steps, reset the target network parameters $\phi^- \leftarrow \phi$.	    
				
			}
		}
	\end{algorithm}

	\section{Hierarchical Distributed Beam Selection}\label{sec:dis}
	In the centralized scheme of Section \ref{sec:cen}, the CPU obtains observations from all BSs and then feeds back the selected beams to the BSs. This process introduces fronthaul link overhead and transmission delay, which may not be acceptable for delay-sensitive users. Moreover, the centralized scheme lacks scalability, has high network training cost and slow convergence speed, due to the large cascaded action spaces. To address these issues, we propose distributed schemes that consider each BS as an agent within a MA system, enabling individual agents to make decisions based on their local observations rather than global state information.
	%% 这里到底节省了多少开销？队列长度不需要汇总（每个基站获得的是一致的），上一时刻的等效信道，有充足时间可以通过fronthual传到CPU。剩下就是beam selection决策要分发给各个基站。

	In this section, based on the hierarchical framework presented in Section~\ref{sec:cen}, we introduce three hierarchical distributed beam selection schemes for the CF system, which partition the system's cascade action space into the effective action space of the BSs. Firstly, we propose a Lyapunov optimization method which is widely utilized for optimizing particular performance metrics while ensuring the stability of network queues. Secondly, we propose a fully distributed scheme using MARL, which further reduces the high beam training overhead required in the hierarchical distributed Lyapunov optimization. Thirdly, based on the fully distributed scheme, we introduce a partially distributed scheme that leverages global state information during the training phase.
	\subsection{Hierarchical Distributed Lyapunov Optimization}
	In~\cite{gatzianas2021traffic}, the authors present a strategy termed Lyapunov-based Coordinated Beamforming (LCB). This strategy takes the stochastic traffic and channel fluctuations into consideration. Remarkably, it is proven that LCB is optimal in terms of throughput and capable to achieve the maximum rate region. The action selection for LCB in time slot $t$ is:
	\begin{equation}\label{equ:lya}
		\mathbf{a}^{\text{LCB}}(t)=\mathop{\arg\max}_{\mathbf{a}(t)\in\mathcal{A}}\sum_{u=1}^{U}q_{u}(t)R_{u}\left(\mathbf{a}(t)\right),
	\end{equation}
	where $R_u(\mathbf{a}(t))$ is the available rate of user $u$ at time slot $t$ when action $\mathbf{a}(t)$ is selected, $\mathcal{A}$ denotes the original action space of the system without any pruning process. Nevertheless, LCB has high computational complexity and needs full CSI across all beams, time slots, and users. Moreover, the computational complexity of LCB is extremely high because it computes the actual rate for each action in the action space.
	
	To overcome these limitations, we integrate the hierarchical framework with the Lyapunov optimization, and propose an algorithm termed HDLO. HDLO is specifically designed for CF beam selection and offers substantial reductions in both computational complexity and beam training overhead when compared to LCB. The beam selection of HDLO in each time slot is carried out through the following process:
	\begin{enumerate}
		\item At the beginning of the long-timescale interval, the candidate beam set ${\mathcal{C}}_{b,u}$ for the pair of BS $b$ and user $u$ is obtained via the wide beam response based CNN prediction. After removing any duplicate beams, i.e. $\mathcal{C}_b=\left\{\mathbf{f}_{b}|\mathbf{f}_{b}\neq\mathbf{f}'_b,\forall\mathbf{f}_b,\mathbf{f}'_b\in\left\{\mathcal{C}_{b,u}\right\}_{u=1}^U\right\}$, the BS $b$ will train the beams in $\mathcal{C}_b$ in each time slot during the long-timescale interval. In addition, the effective action space $\mathcal{A}_{\text{eff},b}$ is obtained according to Eq.~\eqref{equ:eff_b1} and Eq.~\eqref{equ:eff_b}.

		%		\item At the beginning of the long-timescale interval, the effective action space $\mathcal{A}_{\text{eff},b}$ is obtained according to Eq.~\eqref{equ:eff_b1} and Eq.~\eqref{equ:eff_b} using the wide beam response.
		%		Subsequently, the candidate beam set $\mathcal{C}_b$ for the BS $b$, which the BS $b$ will use to train in each time slot during the long-timescale interval, is formed by combining the user's candidate beam set and removing any duplicate beams, i.e. $\mathcal{C}_b=\left\{\mathbf{f}_{b}|\mathbf{f}_{b}\neq\mathbf{f}'_b,\forall\mathbf{f}_b,\mathbf{f}'_b\in\left\{\mathcal{C}_{b,u}\right\}_{u=1}^U\right\}$.

		\item During each time slot, the BS first performs narrow beam training, subsequently evaluating the individual user's rate following each action executed within the effective action space. 
		In distributed schemes, the actual rate of users at a single BS is not available, necessitating the adoption of estimated rates to assess the QoS rendered to users by an individual BS. The distributed design enables the beam selection scheme to be scalable to the number of BSs. Assuming that inter-user interference can be mitigated by subsequent digital processing, and considering only analog combining, the rate provided by BS $b$ to user $u$ is estimated as
		\begin{equation}\label{equ:reward_d}
			\centering
			\mathring{R}_{b,u}=\log_2\left(1+\frac{P\Vert \mathbf{W}_{\text{RF},b}\mathbf{h}_{b,u}\Vert^2}{\Vert \mathbf{W}_{\text{RF},b}\Vert^2}\right).
		\end{equation}
		Significantly, $\mathring{R}_{b,u}$ bears no real physical significance and serves solely as a metric for comparing variations in the QoS delivered by the same BS to different users.
		\item For each BS, based on its effective action space and the estimated rate of each action in the effective action space, the following distributed Lyapunov optimization is performed, and the optimal action is determined as follows
		\begin{equation}\label{equ:hdlo}
			\centering
			\mathbf{a}^{\text{HDLO}}_b(t)=\mathop{\arg\max}_{\mathbf{a}_b(t)\in\mathcal{A}_{\text{eff},b}}\sum_{u=1}^{U}q_{u}(t)\mathring{R}_{b,u}(\mathbf{a}_b(t)).
		\end{equation}
		To be more specific, the rate estimates associated with each action are multiplied by the current user queue length $q_u(t)$ during the time slot, and the resulting products are then summed. 
	\end{enumerate}
	
	Compared to LCB, the proposed HDLO significantly reduces the beam training overhead and computational complexity.
	\begin{itemize}
		
		\item \textbf{Beam training overhead.} The number of narrow beams to be trained in each time slot of HDLO is constrained by $\vert\mathcal{C}_b\vert\leq\min(KU,M)$. When each BS assigns a limited number of candidate beams to each user, it only needs to train a subset of these beams in each time slot. As a result, HDLO reduces the beam training overhead compared to LCB.
		
		\item \textbf{Response delay.} LCB uses the global action $\mathbf{a}=\{\mathbf{a}_b\}_{b=1}^B$ from the original cascaded action space to select beams in the CPU, while HDLO employs the local action $\mathbf{a}_b$ within a greatly smaller local effective action space in the BS. Consequently, the decision-making process for beam selection becomes faster and more efficient.
		
		\item \textbf{Computational complexity.} LCB must compute the actual rate for each global action using Eq.~\eqref{equ:actual_rate}. However, it leads to high computational complexity due to the large cascaded action space and the involved matrix inverse operations. 
		In HDLO, each BS solely computes the estimated rate for the local action by Eq.~\eqref{equ:reward_d}. 
		Notably, this process does not involve any inverse operations, leading to a significant reduction in computational complexity. 
		
	\end{itemize}
	\subsection{Fully Distributed Beam Selection}
	Compared to our proposed centralized RL based WBR-D3QN scheme, HDLO still exhibits a rapid increase in the number of training beams in each time slot, proportionate to the number of candidate beams and users involved. This results in significant beam training overhead for the system, especially when there are a large number of users in the system. 
	To alleviate this burden, we propose a fully distributed MARL approach. This method ensures that the number of training beams in each time slot aligns with the number of users present. 
	In the fully distributed MARL, each BS performs as an agent, independently interacts with the CF system and receives distinct rewards, without sharing local observation information with the other BSs. However, a fully distributed MARL approach may fail to identify the globally optimal action, thereby adversely affecting overall performance.
	
	To address this concern, the proposed hierarchical framework can impose a lower bound on the algorithm's performance by constraining action selection to a specific safe region. We combine the proposed hierarchical framework with a fully distributed (FD) RL algorithm, resulting in the suggested Distributed Double DQN algorithm (D-DDQN) where the RL elements of the agent $b$ can be defined as follows: 
	\subsubsection{\textbf{Action}}
	The action $\mathbf{a}^{\text{FD}}_b(t)$ is denoted as the beam indices assigned by BS $b$ to the users, i.e., $\mathbf{a}^{\text{FD}}_b(t)=\{i_{b,u}(t),u\in\mathbb{U}\}$.
	\subsubsection{\textbf{Local Observation}}
	The variations observed in queue length are a result of joint transmission by multiple BSs in a CF system. In situations where independent decisions are made by individual BSs, the variation in queue length becomes non-stationary for a single BS. Therefore, the local observation $\mathbf{o}^{\text{FD}}_b(t)$ is the local  equivalent CSI, i.e.,
	\begin{equation}\small
		\mathbf{o}^{\text{FD}}_{b}(t)=\left\{\mathbf{W}_{\text{RF},b}(t)\mathbf{h}_{b,u}(t),u\in\mathbb{U}\right\}.
	\end{equation}
	\subsubsection{\textbf{Reward}}
	In D-DDQN, individual rewards are assigned to each agent, and it is imperative to maintain the independence of each agent's reward from the actions undertaken by other agents. Specifically, the reward function is designed to be related to the counterpart agent, and is unrelated to the actions taken by the other agents.
	The instantaneous queue length decreases as the user rate increases in Eq.~\eqref{equ:queue}. From the formulation of the global reward in Eq.~\eqref{equ:cost}, we propose to design the reward for the agent $b$ via utilizing the estimated rate derived from Eq.~\eqref{equ:reward_d} as follows.
	\begin{equation}\label{equ:sep_reward}
		r_b^{\text{FD}}(t)=\sum_{u=1}^{U}\frac{\mathring{R}_{b,u}(\mathbf{a}_b^{\text{FD}}(t))}{\bar{q}_u}.
	\end{equation}
	
	Every agent independently selects actions in the local effective action space by the counterpart local pseudo-sate $\hat{\mathbf{s}}^{\text{FD}}_b(t)=\{\mathbf{o}^{\text{FD}}_{b}(t-1),\mathbf{a}_b^{\text{FD}}(t-1),r_b^{\text{FD}}(t-1),...,\mathbf{o}^{\text{FD}}_{b}(1),\mathbf{a}_b^{\text{FD}}(1),r_b^{\text{FD}}(1)\}.$
	Consequently, after selecting actions, each agent receives rewards from the environment. Owing to space constraints, the algorithmic details of DDQN are omitted and can be found in~\cite{van2016deep}.
	It's noteworthy that the entire network training process is carried out within the BS, without information exchange with the CPU or other BSs. Thus, the signaling overhead is substantially reduced.
	% 限于篇幅，这里DDQN的算法细节就省略了，具体可以参考论文[]

	\subsection{Partially Distributed Beam Selection}
	%	In the fully distributed scheme, the environment is non-stationary \textcolor{red}{because the agents are updating.}
	In the fully distributed scheme, each agent encounters a
	non-stationary environment because of the presence of other
	agents.
	Moreover, each agent only receives the local information, which could lead to a decrease in overall performance. 
	
	To mitigate these issues, we suggest a paradigm called CTDE~\cite{9738819}.
	In the training phase, the CPU aggregates all the local observations to construct a global state. Compared to the independent training of each agent, the CTDE facilitates the training of both the global network in the CPU and the local networks in the agents. This approach yields more favorable training outcomes. The execution of CTDE is also fully distributed. The partially distributed algorithm is identified with MA. 
	
	CTDE schemes based on the actor-critic framework~\cite{huang2019overview} have been extensively studied. The work in~\cite{xu2021multi} achieves joint optimization of beam selection and user scheduling with MA Deep Deterministic Policy Gradient (MADDPG), which has been widely used across various scenarios. In contrast, QMIX is dedicatedly designed for fully cooperative MA environments. Moreover, MADDPG and QMIX respectively take actions in the continuous and discrete action space. Thus, QMIX is more appropriate for our investigated MARL problem which is fully cooperative and has discrete optimization variables.

	The RL elements of the agent $b$ can be defined as follows:
	\subsubsection{\textbf{Action}}
	The action $\mathbf{a}^{\text{MA}}_b(t)$ is the same as $\mathbf{a}^{\text{FD}}_b(t)$, i.e. $\mathbf{a}^{\text{MA}}_b(t)=\mathbf{a}^{\text{FD}}_b(t)$.
	\subsubsection{\textbf{Local Observation}}
	The local observation $\mathbf{o}^{\text{MA}}_b(t)$ is defined as a concatenation of the amount of traffic to be processed and the local equivalent CSI, i.e.,
	\begin{equation}
		\mathbf{o}^{\text{MA}}_{b}(t)=\left\{q_u(t+1),\mathbf{W}_{\text{RF},b}(t)\mathbf{h}_{b,u}(t),u\in\mathbb{U}\right\}.
	\end{equation}
	\subsubsection{\textbf{Reward}}
	Owing to centralized training, we use the $r(t)$ in Eq.~\eqref{equ:cost} as the global reward, instead of the individual reward $r_b^{\text{FD}}(t)$ in Eq.~\eqref{equ:sep_reward}.
	
	The corresponding local pseudo-state is 
		$\hat{\mathbf{s}}^{\text{MA}}_b(t)=\{\mathbf{o}^{\text{MA}}_{b}(t-1),\mathbf{a}_b^{\text{MA}}(t-1),r(t-1),...,\mathbf{o}^{\text{MA}}_{b}(1),\mathbf{a}_b^{\text{MA}}(1),r(1)\}.$ And the gobal pseudo-state is $\hat{\mathbf{s}}^{\text{MA}}(t)=\{\mathbf{o}^{\text{MA}}_{b}(t-1),\mathbf{a}_b^{\text{MA}}(t-1),r(t-1),...,\mathbf{o}^{\text{MA}}_{b}(1),\mathbf{a}_b^{\text{MA}}(1),r(1), \forall b=1,...,B\}.$
	
	Considering the distinct characteristics and network topology inherent in the joint transmission of multi BSs within CF systems, we employ the QMIX algorithm~\cite{rashid2020monotonic} for the problem~\eqref{equ:max} in the MARL environment. 
	Under the CTDE framework, QMIX imposes monotonic constraints on both the global $Q$ value $Q_{\text{tot}}$ and local $Q$ value $Q_b$, i.e.,
	\begin{equation}\label{equ:constraint}
		\frac{\partial Q_{\text{tot}}}{\partial Q_b}\geq0, \forall b\in\mathbb{B}. 
	\end{equation}

	\begin{figure}[!tbp]
		\centering
		\includegraphics[width=0.84\linewidth]{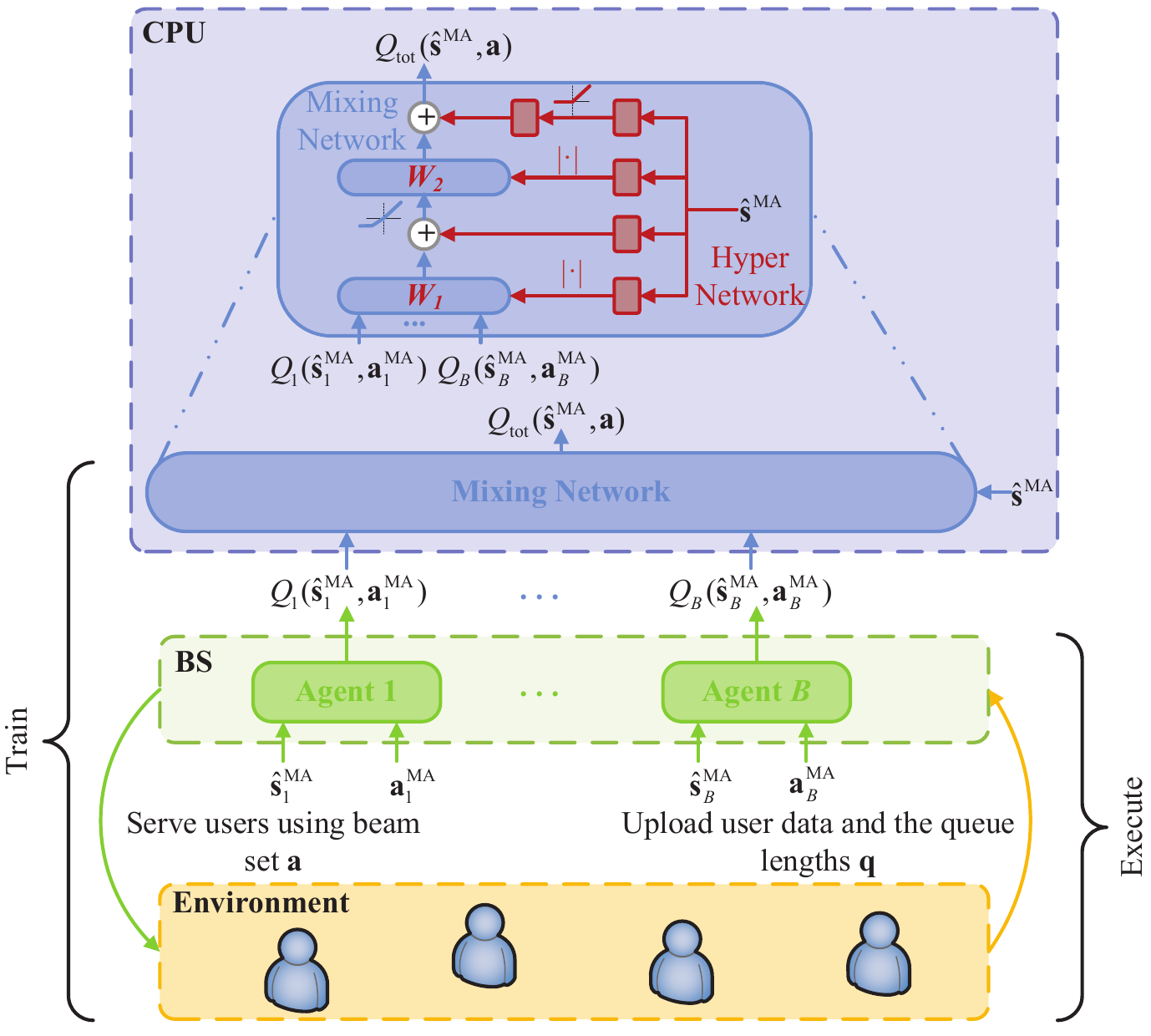}
		\caption{Proposed Partially Distributed Beam Selection Based on QMIX}
		\label{pic:QMIX}
	\end{figure}
	These imposed constraints ensure the monotonicity of both the global and local $Q$ values, making it particularly well-suited for fully cooperative MA tasks. 
	The training process of the proposed QMIX-based partially distributed beam selection scheme (QMIX-PDBS) is illustrated in Fig.~\ref{pic:QMIX}, where each BS is equipped with a DQN which receives local pseudo-state $\hat{\mathbf{s}}^{\text{MA}}_b$ and outputs local $Q$ value $Q_b$. In the CPU, a mixing network aggregates local $Q$ values from all agents and outputs the global $Q$ value $Q_{\text{tot}}$. The mixing network's weights and bias are provided from a learnable hypernetwork which takes the global pseudo-state $\hat{\mathbf{s}}^{\text{MA}}$ as the input. Notably, to satisfy the constraint~\eqref{equ:constraint}, the outputs of the hypernetwork must be non-negative, which is realized by a ReLU or an absolute activation function.
	With the assistance of $Q_{\text{tot}}$, the QMIX network including local DQNs at multiple BSs and the mixing network in the CPU is trained in an end-to-end manner to minimize the following loss
	\begin{equation}\label{equ:loss}
		\begin{split}
			L^{\text{MA}}(\theta)=\sum_{n=1}^{N}\Big(r_n+\gamma\max_{\mathbf{a}_n'\in\mathcal{A}_{\text{eff}}}Q^-_{\text{tot}}(\hat{\mathbf{s}}^{\text{MA}'}_n,\mathbf{a}_n';\theta^-)\\
			-Q_{\text{tot}}(\hat{\mathbf{s}}^{\text{MA}}_n,\mathbf{a}_n;\theta)\Big)^2,
		\end{split}
	\end{equation}
	where $N$ denotes the batch size sampled from the replay buffer, $Q^-_{\text{tot}}$ denotes the target network,  $\hat{\mathbf{s}}^{\text{MA}'}$ denotes the next global pseudo-state after taking action $\mathbf{a}=\{\mathbf{a}_b^{\text{MA}}\}_{b=1}^B$ in the current gloabl pseudo-state $\hat{\mathbf{s}}^{\text{MA}}$.	
	The training phase of the proposed QMIX-PDBS scheme is summarized in Algorithm~\ref{alg:train}.
	\begin{algorithm}[ht!]
		\caption{Proposed QMIX-PDBS Scheme (Training Phase)}
		\label{alg:train}
		All BSs receive wide beam probing and the CPU constructs the effective action space for each BS by the prediction of CNN.
		
		Initialize the replay buffer $\mathcal{D}$, the local networks with random parameters $\phi_b, b\in\mathbb{B}$, the mixing network with random parameters $\xi$ and the target network with parameters $\xi^-=\xi$.
		
		\BlankLine
		\For{$episode=1:total\_episode$}{
			Reset the environment, and randomly initialize queues for all users.
			
			\For{$t=1:T$}{
				\eIf{$t=1$}{
					Each BS $b\in\mathbb{B}$ randomly chooses the beam set $\mathbf{a}_b^{\text{MA}}(t)$.
				}{Each BS $b\in\mathbb{B}$ takes action $\mathbf{a}_{b}^{\text{MA}}(t)$ based on its local pesudo-state $\hat{\mathbf{s}}_b^{\text{MA}}(t)$.}
				
				Each BS $b\in\mathbb{B}$ conducts the beam selection and data transmission and updates its local pseudo-state $\hat{\mathbf{s}}_b^{\text{MA}}(t+1)$ using the local equivalent CSI $\mathbf{W}_{\text{RF},b}(t)\mathbf{h}_{b,u}(t)$ and the queue length $q_u(t+1), \forall u\in\mathbb{U}$.
				
				Each BS $b\in\mathbb{B}$ uploads the local pesudo-state $\hat{\mathbf{s}}_b^{\text{MA}}(t+1)$ and the local $Q$ value $Q_b$ to the CPU. 
				
				The CPU stores $<\hat{\mathbf{s}}^{\text{MA}}(t),\mathbf{a}^{\text{MA}}(t),r(t),$ $\hat{\mathbf{s}}^{\text{MA}'}(t+1)>$ in the replay buffer $\mathcal{D}$.
				
				The CPU updates the mixing network and the local $Q$ networks according to Eq.~\eqref{equ:loss}.
				
				After $N_c$ steps, reset the target network parameters $\xi^- \leftarrow \xi$.
				
			}
		}
	\end{algorithm}
	
	As shown in Fig.~\ref{pic:QMIX}, in the execution phase, each agent interacts with the environment independently, receives the local pseudo-state $\hat{\mathbf{s}}_b^{\text{MA}}$, and selects the set of beams $\mathbf{a}_b^{\text{MA}}$ from the effective action space to serve the user based on the local $Q$ network. Crucially, this process does not necessitate communication with the CPU for passing the local pseudo-state $\hat{\mathbf{s}}_b^{\text{MA}}$.

	%	In summary, the partially distributed algorithm exhibits the capacity to leverage global information to improve the global performance in the training phase. Simultaneously, during the execution phase, the agents no longer necessitate interactions with the BS, thus effectively reducing signaling exchange overhead and system response delay.

	\section{Numerical Results}\label{sec:sim}
	\subsection{Simulation Setup}
	Considering a region of 150m$\times$150m composed of 3 BSs and 4 or 8 randomly distributed users. The BSs are separated by at least 75m and the users are separated by at least 10m. Each BS is equipped with an 8$\times$4 UPA, and each user has a single antenna. The heights of BSs and users respectively are 6m and 2m. The users' positions change across long-timescale intervals and are fixed within a long-timescale interval. Thus, the large-scale fading, azimuth and elevation angles remain constant during a long-timescale interval. The small-scale fading obeys the first-order autoregressive process $\beta_{b,u,l}(t)=\rho\beta_{b,u,l}(t-1)+\sqrt{1-\rho^2}n_{b,u,l}(t)$ where $\rho$ is the correlation coefficient. Following the 3GPP channel, the channel path is either blocked, line-of-sight (LoS), or non-LoS (NLoS) \cite{38901}. The carrier frequency is $f_\text{c}=28$GHz, and the bandwidth is $ W=100 $MHz. The user transmit power is 0.2W, and the BS receiver noise power is $p_{\text{n}} = 3.18\times 10^{-12}$W (i.e., noise temperature $T_0=290$K, noise factor $\sigma=9$dB). Each user has a unique service type, traffic pattern and delay requirement. 
	An episode includes a long-timescale interval consisting of 100 time slots. The parameter settings are listed in Table~\ref{tab:env}.
	
	\begin{table}[htb]
		\caption{Parameter Setup}
		\begin{center}
			\resizebox{\linewidth}{23mm}{
				\begin{tabular}{l|l}
					\hline
					\textbf{Parameters}&\textbf{Values} \\ \hline
					Package shape $\kappa$, threshold $\chi_{\min}$&$6$, $1$bit\\
					\hline
					4-user package arrival rate $\lambda_u,u\in\mathbb{U}$&$[4.5,5,5.5,6]$ packets/slot\\ \hline
					4-user delay requirement $\bar{q}_u,u\in\mathbb{U}$ &$[9,10,11,12]$ bits\\\hline
					4-user delay limit $\breve{q}_u,u\in\mathbb{U}$ &$[18,20,22,24]$ bits \\\hline
					8-user package arrival rate $\lambda_u,u\in\mathbb{U}$&$[5.5,5,5.5,4.5,6,5,6,4.5]$ packets/slot\\ \hline
					8-user delay requirement $\bar{q}_u,u\in\mathbb{U}$ &$[11,10,11,9,12,10,12,9]$ bits\\\hline
					8-user delay limit $\breve{q}_u,u\in\mathbb{U}$ &$[22,20,22,18,24,20,24,18]$ bits\\\hline
					Path number $L$&6\\\hline
					Correlation coefficient $\rho$&0.91\\  \hline
					Time slot duration $\tau$& 1 ms\\ \hline
					Symbol duration $\tau_\text{c}$ & 5 $ \mu $s\\ \hline
					Weighting factor $\delta$ &10 \\ \hline
				\end{tabular}
				\label{tab:env}
			}
		\end{center}
	\end{table}
	% 表中单位为什么是packets？？？

	\begin{figure*}[t]
		\centering
		\subfloat[Sample 1]{\includegraphics[width=0.4\linewidth]{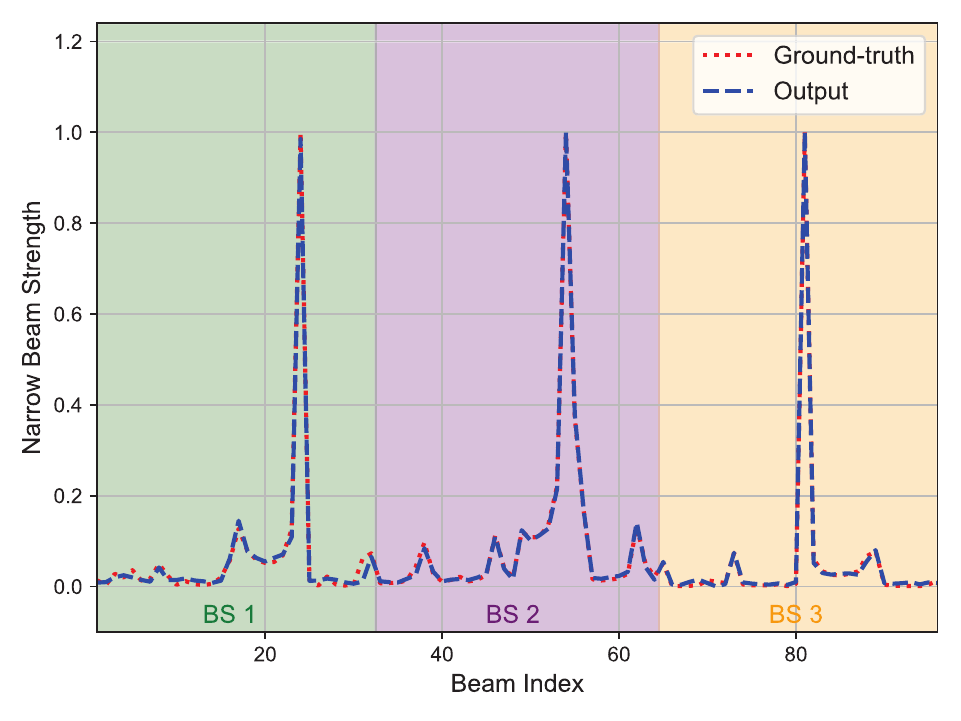}}
		\subfloat[Sample 2]{\includegraphics[width=0.4\linewidth]{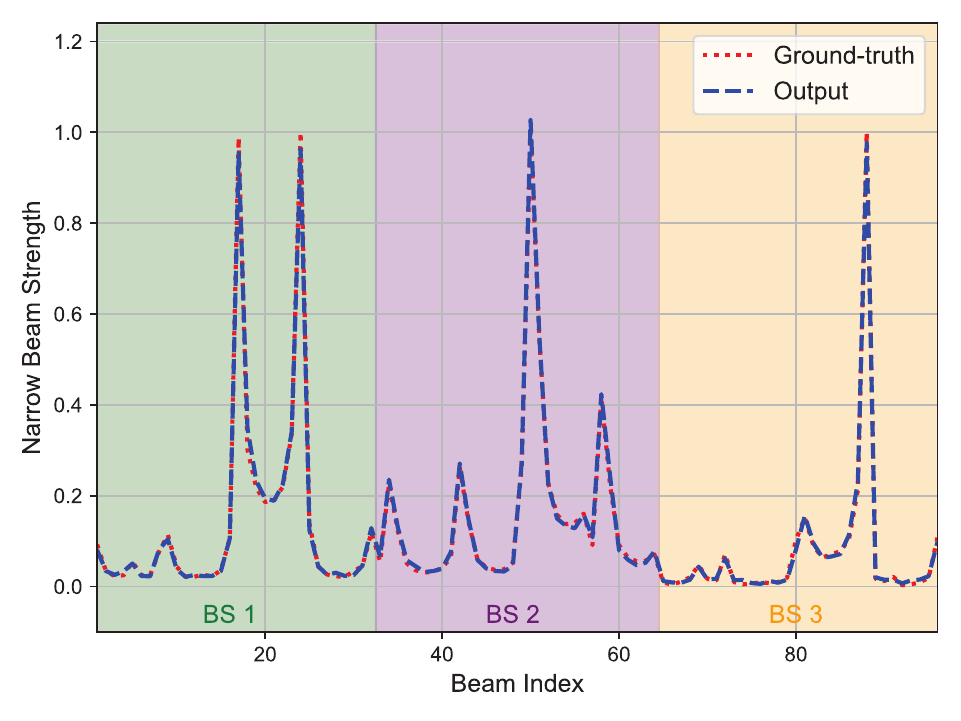}}
		\caption{CNN prediction results.}
		\label{pic:predict}
		% 设置子图标题的字体大小为footnotesize
	\end{figure*}
	\subsection{Prediction of Narrow Beam Power Profile}
	In the simulations, 8 antennas are activated to form the wide beam using the DFT codebook. The input dimension of the CNN is $3\times 8=24$ and the output dimension is $3\times 32=96$. 
	The CNN comprises three convolutional layers, each containing 16 kernels with a size of $2\times2$, and three cascaded fully connected layers, each consisting of 256 neurons. 
	%输入维度是多少 宽波束是什么？
	The rectified linear unit (ReLU) is used as the activation function, the Stochastic Gradient Descent with Momentum (SGDM) optimizer is implemented, and a learning rate schedule with gradual reduction is adopted.

	%这个图太小了 可以考虑占满一整行
	
	To construct the data set, each BS performs the wide beam probing followed by the narrow beam probing in a single time slot, and the beam strengths obtained from the two responses are uploaded to the CPU to construct the data set for CNN training. The dataset comprises 30,000 samples in total, partitioned into three subsets for training, validation, and testing with sizes of 21000, 4500, and 4500, respectively. We randomly select 2 samples from the test set and compare the CNN output with the actual narrow beam strength, as shown in Fig.~\ref{pic:predict}. It is worth noting that the beams indexed within the ranges 1 to 32, 33 to 64, and 65 to 96 respectively are attributed to BS 1, BS 2, and BS 3. According to the predicted performance on the test set, the CNN can precisely learn the relationship between the wide beam strengths and the narrow beam strengths of multiple BSs well, and it can accurately predict multiple beams with relatively high strength. The experiments tested the performance of the trained CNN. The accuracy of the predicted strongest 2 beams is $89.74\%$. The accuracy that the predicted strongest 2 beams containing the actual strongest beam is $99.73\%$. We select the strongest 2 beams for each user to construct the system cascaded action space for the centralized scheme. While the distributed schemes have a significantly smaller action space compared to the centralized scheme. In order to improve the delay satisfactory rate, the strongest 3 beams are selected to construct the BS effective action space for the distributed schemes.

	\subsection{Delay Satisfaction Rate of Centralized Algorithms}
	\begin{table}[tbp]
		\caption{WBR-D3QN Hyper Parameters}
		\begin{center}
			\begin{tabular}{l|l}
				\hline
				\textbf{Parameters}&\textbf{Values}\\ \hline
				Advantage network &[128,256,256,256,1]\\ \hline
				Value network &[128,256,256,256,384]\\ \hline
				Learning rate &0.005\\ \hline
				Batch size &256\\ \hline
				Buffer Size &10000\\ \hline
				Activation function &ReLU\\ \hline
				Discount factor $\gamma$&0.99\\ \hline
				Target network update interval $N_c$&4 steps\\ \hline
			\end{tabular}
			\label{tab:wbr}
		\end{center}
	\end{table}
	Simulations are conducted to compare the delay satisfaction rates between the proposed centralized scheme and the existing traffic-aware beam selection schemes.
	The hyper-parameters of the proposed WBR-D3QN scheme are shown in Table~\ref{tab:wbr}. 
	%%表二中的一连串数字代表什么意思？
	The value network and advantage network of WBR-D3QN share the first 3 network layers for feature extraction. In order to validate the effectiveness and indispensability of the proposed framework, two algorithms for comparison are given as follows: 
	\subsubsection{LBS}
	To overcome the high computational complexity of LCB, a heuristic algorithm (LBS) is proposed in~\cite{gatzianas2021traffic} that prunes the set of available beams for each user to the $K$ strongest beams after beam training and then applies the same beam selection framework as LCB.
	The performance of LBS is shown to be comparable to LCB in the ideal case of no beam conflicts and no interference between users. At each time slot, LBS selects actions based on the CSI over partial beams and the queue lengths.
	\begin{equation}
		\mathbf{a}^{\text{LBS}}(t)=\mathop{\arg\max}_{\mathbf{a}(t)\in\mathcal{A}^{\text{LBS}}}\sum_{u=1}^{U}q_{u}(t)R_{u}(\mathbf{a}(t)),
	\end{equation}
	where $\mathcal{A}^{\text{LBS}}\subseteq \mathcal{A}$ denotes the system action space of LBS after pruning. LBS significantly decreases the computational complexity of LCB but does not reduce the beam training overhead of LCB.
	
	\subsubsection{SBA-D3QN}Accordingly, the joint design without wide beam response's help results in an enormous action space with $4.6501\times10^{13}$ actions, which is infeasible. We assume that the indices of the strongest beams from the BS to each user are known, and the action space can be pruned in advance. The continuous 3 beams are selected as the action space for each user, where the second beam is the strongest beam. In the experimental setting, it is observed that the action space of SBA-D3QN, utilizing the pruning method mentioned earlier, is approximately an order of magnitude greater than the action space employed by WBR-D3QN proposed.
	
	\begin{table}[!tbp]
		\caption{Comparison of different centralized algorithms\tnote{1}}
		\centering
		\label{tab:cent}
		\setlength{\tabcolsep}{3pt}{
			\begin{tabular}{c|c|c|c}
				\hline
				&\textbf{WBR-D3QN}&\textbf{LBS}&\textbf{SBA-D3QN}\\ \hline
				Beam training overhead &$U\times \tau_\text{c}$ & $M\times \tau_\text{c}$&$U\times \tau_\text{c}$ \\ \hline
				Computational complexity &$\mathcal{O}(1)$&$\mathcal{O}(\vert\mathcal{A}_{\text{LBS}}\vert)$&$\mathcal{O}(1)$\\
				\hline
				System delay satisfaction rate&$\textbf{70.71\%}$&$21.08\%$&$22.07\%$\\\hline
			\end{tabular}
		}
		\begin{tablenotes}
			\footnotesize
			\item[1] $\mathcal{O}(1)$ denotes one ZF beamforming.
		\end{tablenotes}
	\end{table}
	\begin{figure}[!t]
		\centering	
		\includegraphics[width=0.9\linewidth]{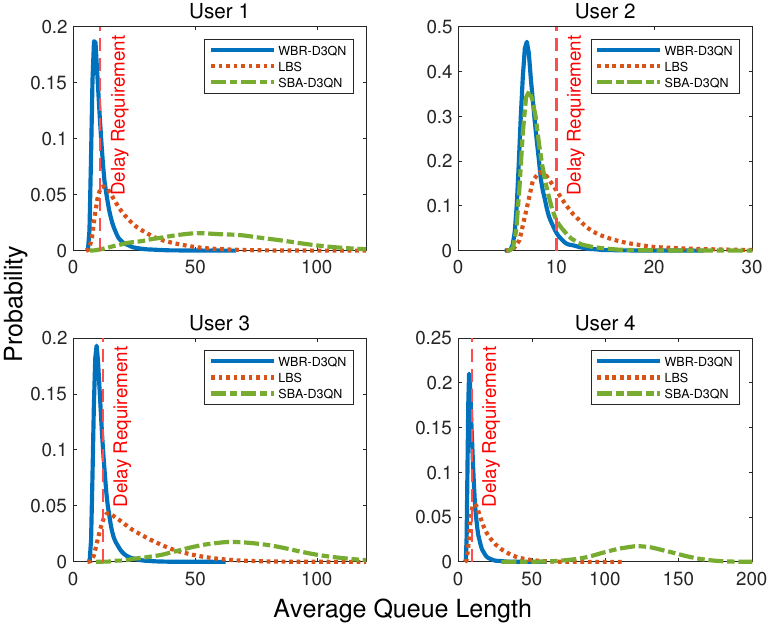}
		\caption{Distribution of different users' average queue lengths for centralized algorithms.}
		\label{pic:queue_centr}
	\end{figure}
	
	\begin{figure}[t]
		\centering
		\includegraphics[width=0.85\linewidth]{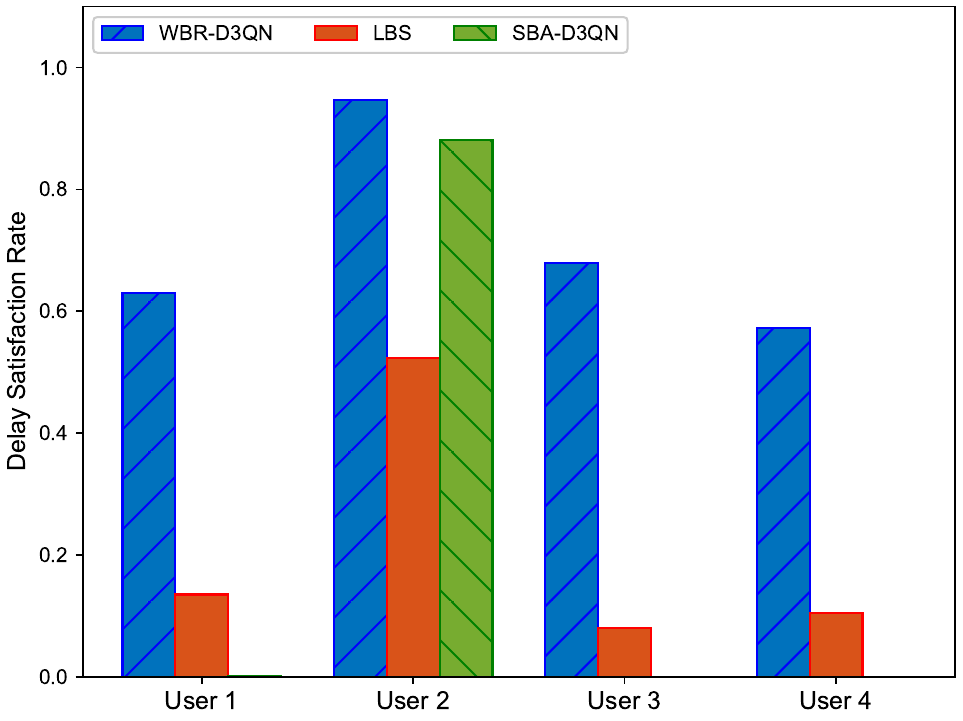}
		\caption{Delay satisfaction rate versus different users.}
		\label{pic:delay_centr}
	\end{figure}
	
	Under the same environmental conditions, we conducted 20,000 simulation experiments to evaluate and compare the average queue length distributions of the centralized algorithms. The corresponding results are represented in Fig.~\ref{pic:queue_centr}. It is imperative for a high-performance algorithm to maintain low average delay and minimal delay jitter to meet delay requirements and ensure satisfactory user QoS. In essence, the algorithm should achieve both a low average queue length and a high concentration of queue length values around that average, thereby promoting a more focused and stable overall distribution. As illustrated in Fig.~\ref{pic:queue_centr}, it is evident that WBR-D3QN outperforms other algorithms in providing a better service to users.
	
	The delay satisfaction rates of different users are shown in Fig.~\ref{pic:delay_centr}. The performance of WBR-D3QN is superior to all the baselines. SBA-D3QN tends to provide more resources to the user with good communication quality, while the delay satisfaction rate for the users with low communication quality is extremely low. Meanwhile, LBS cannot effectively meet user delay requirements.
	In particular, SBA-D3QN converges to a poor suboptimal solution which leads to the failure to satisfy the delay requirements of the users. On the other hand, LBS performs exhaustive beam sweeping of all users at each time slot, and thus the uploading traffic is sharply reduced.
	Furthermore, Table~\ref{tab:cent} presents a comparative analysis of beam training overhead, computational complexity and the system delay satisfaction rate (i.e., the average of all users' delay satisfaction rates) of different algorithms. Note that LBS computes the actual rate for all actions within the action space and selects the action with the largest weighted sum as the final action, which also causes a non-negligible delay before uploading traffic.
	\begin{table}[htb!]
		\caption{QMIX-PDBS Hyper Parameters}
		\begin{center}
			\begin{tabular}{l|l}
				\hline
				\textbf{Parameters}&\textbf{Vaules}\\ \hline
				Mixing network &[256,256]\\ \hline
				Local $Q$ network &[128,128]\\ \hline
				Learning rate &0.01\\ \hline
				Batch size &256\\ \hline
				Replay buffer size &50000\\ \hline
				Activation function &ReLU\\ \hline
				Discount factor $\gamma$&0.99\\ \hline
				Target network update interval $N_c$&2 episodes\\ \hline
			\end{tabular}
			\label{tab:qmix}
		\end{center}
	\end{table}
	
	\subsection{Delay Satisfaction Rate of Distributed Algorithms}
	The effectiveness and scalability of the proposed distributed schemes are substantiated through simulations.
	Within the distributed context, an evaluation of 
	QMIX-PDBS, MADDPG, HDLO, and D-DDQN's performance is conducted in both the 4-user and 8-user scenarios. Notably, the 8-user scenario entails the presence of two users per service type. The hyper parameters of QMIX-PDBS are presented in Table~\ref{tab:qmix}. 
	Meanwhile, in the MADDPG framework, each BS is equipped with an individual actor network that takes local observations to output $Q$ values associated with each action in the effective action space. The optimal action is determined by selecting the one associated with the highest $Q$ value in the execution phase. Concurrently, the CPU maintains a critic network for each BS which takes the system state and one-hot vectors representing the output action of all actor networks as input and yields the global state-action value as output.
	In contrast, in the D-DDQN scheme, each BS is equipped with a DQN that does not communicate with the CPU.
	\begin{table*}[!t]
		\caption{Comparison of different distributed algorithms\tnote{1}}
		\centering
		\begin{tabular}{c|c|c|c|c}
			\hline
			&\textbf{QMIX-PDBS}&\textbf{MADDPG}&\textbf{HDLO}&\textbf{D-DDQN}\\ \hline
			Beam training overhead &$U\times \tau_\text{c}$ & $U\times \tau_\text{c}$&$K\times U\times \tau_\text{c}$&$U\times \tau_\text{c}$ \\ \hline
			Computational complexity &$\mathcal{O}(1)$&$\mathcal{O}(1)$&$\mathcal{O}(1)$&$\mathcal{O}(1)$\\
			\hline
			4-user system delay satisfaction rate&$\textbf{67.66\%}$&$50.87\%$&$61.56\%$&$64.15\%$\\\hline
			8-user system delay satisfaction rate&$\textbf{58.33\%}$&$38.27\%$&$31.47\%$&$57.17\%$\\\hline
		\end{tabular}
		\label{tab:dis}
		\begin{tablenotes}
			\footnotesize
			\item[1] $\mathcal{O}(1)$ denotes one ZF beamforming.
		\end{tablenotes}
	\end{table*}
	\begin{figure}[!t]
		\centering
		\includegraphics[width=0.9\linewidth]{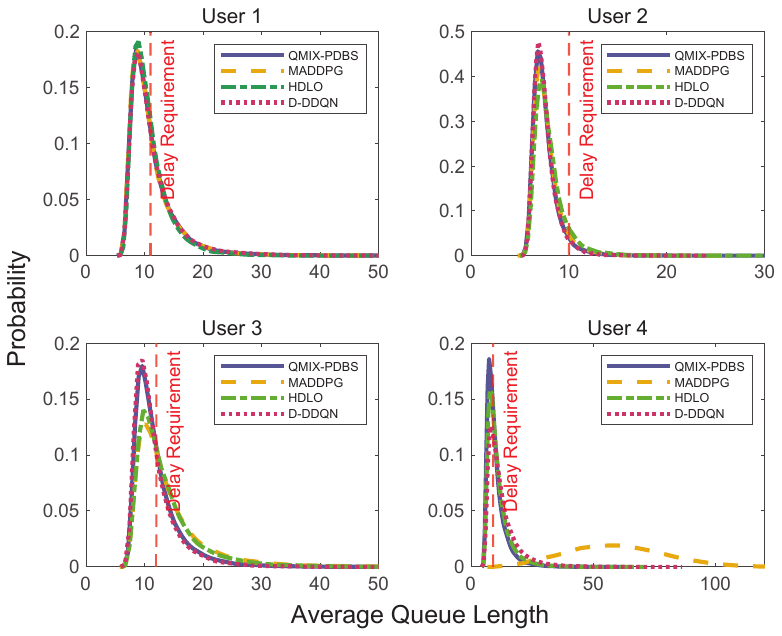}
		\caption{4-user distribution of average queue lengths for distributed algorithms.}
		\label{pic:queue_dis}
	\end{figure}
	
	Fig.~\ref{pic:queue_dis} presents the results of the average queue length distribution for the distributed algorithms in the 4-user scenario. Notably, all algorithms demonstrate stable QoS for users with good communication quality. However, for users 2 and 3, QMIX-PDBS and HDLO show higher peaks compared to the other algorithms, indicating a more concentrated queue length distribution. Conversely, MADDPG exhibits inferior QoS for user 4, who operates in a relatively poor communication environment.
	\begin{figure}[!t]
		\centering
		\includegraphics[width=0.85\linewidth]{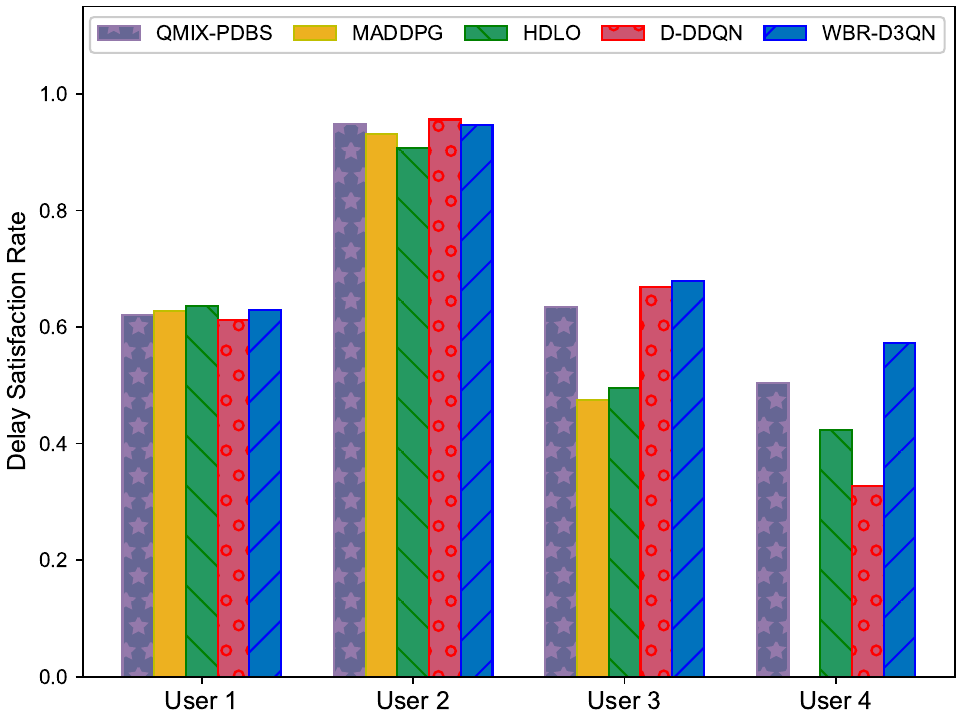}
		\caption{Delay satisfaction rate versus 4 users.}
		\label{pic:delay_dis}
	\end{figure}
	\begin{figure}[!t]
		\centering
		\includegraphics[width=0.85\linewidth]{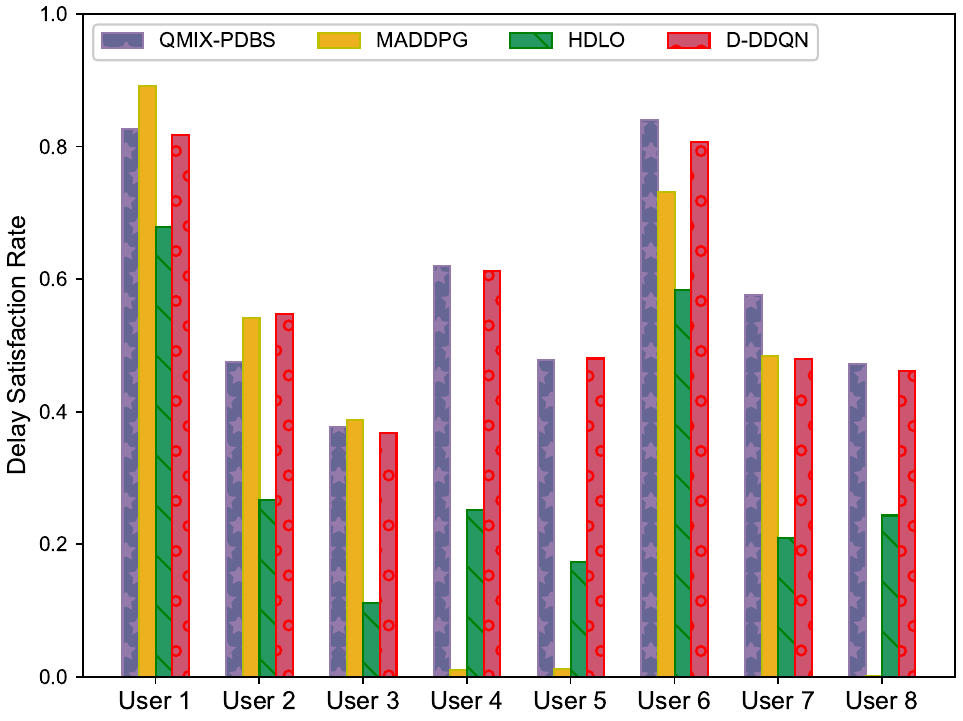}
		\caption{Delay satisfaction rate versus 8 users.}
		\label{pic:delay_dis_ue8}
	\end{figure}
	In contrast to the centralized scheme, the distributed schemes demonstrate significantly faster response times during the execution phase as they do not require communication with the CPU, which effectively reduces system delay.  
	Fig.~\ref{pic:delay_dis} presents the delay satisfaction rate of the distributed algorithms within the context of a 4-user scenario, with QMIX-PDBS achieving the closest performance to the centralized algorithm WBR-D3QN. In contrast, MADDPG exhibits deficiencies, resulting in unstable performance across different users. Moreover, D-DDQN experiences a marginal decline in overall performance compared to QMIX-PDBS, which is attributed to its inability to access global state information during the training phase. Furthermore, it is observed that HDLO is inclined to select actions associated with a more favorable delay satisfaction rate. Nevertheless, the high beam training overhead reduces its effective achievable rate.
	
	A comprehensive comparison of beam training overhead, computational complexity, and system delay satisfaction rate for the distributed algorithms is presented in Table~\ref{tab:dis}. For our experiment, the value $K=3$ is set, which represents the number of candidate beams assigned to each user by the BS. It is worth noting that, in scenarios with low beam conflict probabilities, setting $K$ to 2 can expedite the convergence of the RL method and enhance the performance of HDLO.
	\begin{figure*}[t]
		\centering
		\includegraphics[width=0.9\linewidth]{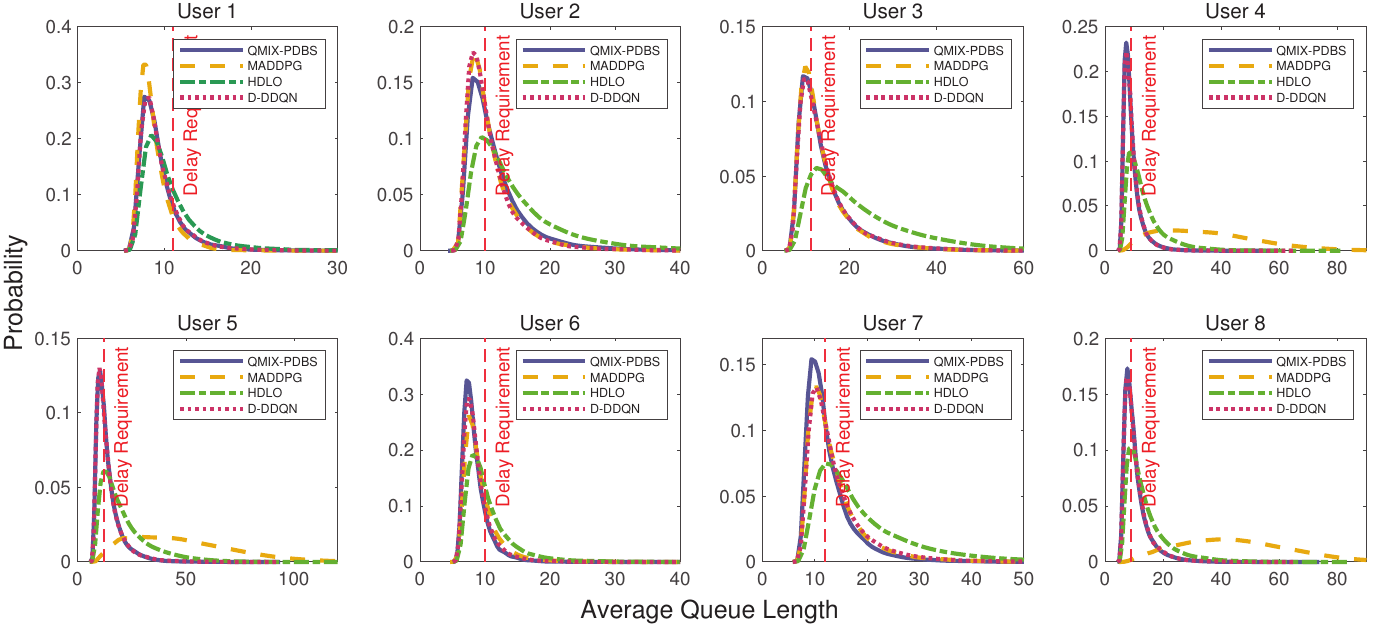}
		\caption{8-user distribution of average queue lengths for distributed algorithms.}
		\label{pic:queue_dis_ue8}
	\end{figure*}
	
	Fig.~\ref{pic:delay_dis_ue8} and Fig.~\ref{pic:queue_dis_ue8} respectively depict the delay satisfaction rate and average queue length distribution of the distributed algorithms in the 8-user scenario. Notably, the cascaded action space of the centralized scheme WBR-D3QN proves to be excessively large, rendering it challenging to train and leading to convergence issues. Consequently, this approach is not feasible for scenarios involving a substantial number of served users.
	In comparison to the 4-user scenario, the system delay satisfaction rate of each distributed algorithm in the 8-user scenario experiences a notable decline due to heightened inter-user interference and limited system resources. Particularly, HDLO exhibits the most substantial decrease in the system delay satisfaction rate, primarily attributed to a significant increase in beam training overhead.
	In contrast to MADDPG, which sacrifices QoS for certain users to enhance the QoS of others within the system, HDLO still achieves a more balanced performance across all users. However, it results in a slight increase in average delay for each user but significantly worse performance in delay satisfaction rate due to its much larger beam training overhead, compared to the top-performing QMIX-PDBS. Furthermore, D-DDQN, due to its lack of interactions between the BS and the CPU, as well as between individual BSs, its system delay satisfaction rate is slightly lower than that of QMIX-PDBS.

	\section{Conclusion}\label{sec:con}
	Considering joint transmission in CF-mMIMO systems, we investigated the traffic-aware beam selection problem with concerns on QoS, and proposed a hierarchical beam selection scheme. The scheme operated on a dual timescale. Firstly, in the long-timescale interval, the CPU adopted a CNN to predict the narrow beam profile with aggregated received wide beams. Based on the CNN prediction, the beam space for traffic was pruned for subsequent beam selection. Secondly, in the short-timescale, the D3QN is used to select the beams for delay satisfaction rate maximizing which is modeled as a POMDP. By modeling the CF-mMIMO system as an MA system, this centralized scheme is further decoupled as a distributed scheme. As a result, hierarchical distributed Lyapunov optimization, fully distributed RL, and QMIX-based CTDE of RL were proposed based on the hierarchical framework. Simulation results showed that the proposed schemes significantly reduced the beam training overhead and computational complexity and also better satisfied the user-specific delay requirement, compared to the baselines.

	\bibliographystyle{IEEEtran}
	\bibliography{wcy_97.bib}

	\vfill
	
\end{document}